\documentclass[vecarrow]{svmult}
\usepackage{graphicx}
\usepackage{maple2e}
\DefineParaStyle{Maple Output}
\DefineCharStyle{2D Math}
\DefineCharStyle{2D Output}

\usepackage{amssymb}
\usepackage{amsmath}
\usepackage{ifthen}
\usepackage{version}
\usepackage{subeqnar}
\usepackage{epsfig}
\usepackage{citesort}
\newcommand{\plist}[1]{\ifthenelse{\equal{\rein}{#1}}{\includeversion{thisone}}{\excludeversion{thisone}} \begin{thisone} \begin{Verbatim}}
\newcommand{\rein}{a}
\newcommand{\proginput}[2]{
\RecustomVerbatimEnvironment{Verbatim}{Verbatim}
  {gobble=0,commentchar=£,numbers=left,numbersep=3pt,frame=single,xleftmargin=0.4cm,xrightmargin=0.1cm,firstnumber=last}
\renewcommand{\rein}{#2}{\scriptsize \input{#1}}
}
\newcommand{\newprog}{\setcounter{FancyVerbLine}{0}}
\newcommand{\eqeqref}[2]{(\ref{#1},\ref{#2})}
\newcommand{\eqeqeqref}[3]{(\ref{#1},\ref{#2},\ref{#3})}
\newcommand{\eqtoref}[2]{(\ref{#1}-\ref{#2})}

\newcommand{\li}[1]{\fbox{#1}}
\usepackage[baw]{fvrb-ex}
\RecustomVerbatimEnvironment{Verbatim}{Verbatim}
  {gobble=0,commentchar=£,numbers=left,numbersep=3pt,frame=single,xleftmargin=0.5cm,xrightmargin=0.2cm}

\begin{document}
%\mainmatter%%%%%%%%%%%%%%%%%%%%%%%%%%%%%%%%%%%%%%%%%%%%%%%%%%%%%%%

%\documentclass[runningheads]{STUFF/svmult}
%\input{STUFF/GGINPUT}

\bibliographystyle{unsrt}

\setcounter{page}{131}
\title*{Kinetic Integrals in the Kinetic Theory\\ of dissipative gases}
\titlerunning{Kinetic Integrals in the theory of dissipative gases}
\toctitle{Kinetic integrals in the kinetic theory of dissipative gases}

\author{Thorsten P\"oschel \and Nikolai V. Brilliantov}
\tocauthor{Thorsten P\"oschel, Nikolai Brilliantov}
\authorrunning{T. P\"oschel \and N. V. Brilliantov}

\institute{Institut f\"ur Biochemie, Charit\'e, Humboldt-Universit\"at zu Berlin,\\ Monbijoustra{\ss}e 2, D-10117 Berlin, Germany}

\maketitle
\vspace*{-6cm} In: T. P\"oschel, N. Brilliantov (eds.) ``Granular Gas Dynamics'',\\ Lecture Notes in Physics, Vol. 624, Springer (Berlin, 2003), p. 131-162
\vspace*{5cm}

\label{page:Poeschel}
\begin{abstract}
The kinetic theory of gases, including Granular Gases, is based on the Boltzmann equation. Many properties of the gas, from the characteristics of the velocity distribution function to the transport coefficients may be expressed in terms of functions of the collision integral which we call kinetic integrals. Although the evaluation of these functions is conceptually straightforward, technically it is frequently rather cumbersome. We report here a method for the analytical evaluation of kinetic integrals using computer algebra. We apply this method for the computation of some properties of Granular Gases, ranging  from the moments of the velocity distribution function to the transport coefficients. For their technical complexity most of these quantities cannot be computed manually. 
\end{abstract}

\section{Introduction} 
\label{sec:Intro}

The collision of elastic hard spheres of initial velocities $\vec{v}_1$ and $\vec{v}_2$ is described by the collision law 
\begin{equation}
\label{Vel:collrulreelast}
    \vec{v}^{\,\prime}_1=\vec{v}_1-\left(\vec{v}_{12}\cdot \vec{e}\,\right)\vec{e}\,,\qquad\qquad
    \vec{v}^{\,\prime}_2=\vec{v}_2+\left(\vec{v}_{12}\cdot \vec{e}\,\right)\vec{e}
\end{equation} 
where $\vec{v}^{\,\prime}_{1/2}$ denote their velocities after the collision, $\vec{v}_{12}\equiv\vec{v}_1-\vec{v}_2$ and $\vec{e}\equiv \vec{r}_{12}/r_{12}$ with $\vec{r}_{12} \equiv \vec{r}_1-\vec{r}_2$ at the instant of the collision. For dissipatively colliding particles the coefficient of restitution $\varepsilon \in (0,1)$ characterizes the loss of relative normal velocity of the particles; $\varepsilon=1$ represents elastic collisions, $\varepsilon=0$ stands for sticky collisions where the relative velocity after a collision drops to zero. Consequently, the collision rule for dissipatively colliding particles reads
\begin{equation}
\label{Vel:15a}
    \vec{v}^{\,\prime}_1=\vec{v}_1-\frac{1+\varepsilon}{2}\left(\vec{v}_{12}\cdot \vec{e}\right)\vec{e}\,,\qquad\qquad
    \vec{v}^{\,\prime}_2=\vec{v}_2+\frac{1+\varepsilon}{2}\left(\vec{v}_{12}\cdot \vec{e}\right)\vec{e} \,.  
\end{equation} 
The kinetic theory uses the concepts of direct and inverse collisions:
In a {\em direct collision}, as described above, the particle velocities change by $\left\{\vec{v}_1,\vec{v}_2\right\}\rightarrow \left\{\vec{v}^{\,\prime}_1,\vec{v}_2^{\,\prime}\right\}$. An {\em inverse collision} is a collision which results in the velocities $\vec{v}_{1/2}$, i.e. $\left\{\vec{v}^{\,\prime\prime}_1,\vec{v}_2^{\,\prime\prime}\right\} \rightarrow \left\{\vec{v}_1,\vec{v}_2\right\}$.

The starting point of the kinetic theory of gases of elastic particles \cite{ChapmanCowling:1970,FerzigerKaper:1972} as well as dissipatively colliding particles (Granular Gases) \cite{PoeschelLuding:2001} is the Boltzmann equation. It describes the evolution of the distribution function $f\left(\vec{r},\vec{v}, t\right)$ of the number of gas particles at time $t$ in the infinitesimal phase-space volume $d\vec{r}, \, d\vec{v}$ located at $\vec{r}, \, \vec{v}$. For dissipative gases it reads (e.g. \cite{PoeschelLuding:2001,BrilliantovPoeschelStability:2000,BrilliantovPoeschel:2000visc})
\begin{equation}
\begin{split} 
\left( \frac{\partial}{\partial t} + \vec{v}_1 \cdot \vec{\nabla} \right) f\left(\vec{v}_1,t\right)
& =\sigma^{d-1} \int d \vec{v}_2 \int d\vec{e} \, 
\Theta\left(-\vec{v}_{12} \cdot \vec{e}\right) \left|\vec{v}_{12} \cdot \vec{e}\right| \times\\
&~~~~~~~~~~~~~\times \left[\chi f\left(\vec{v}^{\,\prime\prime}_1,t\right)f\left(\vec{v}^{\,\prime\prime}_2,t\right)-
f\left(\vec{v}_1,t\right)f\left(\vec{v}_2,t\right) \right]\\
&\equiv I\left(f,f\right)\,,
\label{collint1}
\end{split}
\end{equation} 
which defines the collision integral $I\left(f,f\right)$. Here $\sigma$ is the particle diameter and $d$ is the dimension of the system. The integration is performed over the velocity $\vec{v}_2$ of the particle $2$ of a colliding pair $\{1-2\}$ and over the unit vector $\vec{e}$. $\Theta(x)$ denotes the Heaviside step-function. The Boltzmann equation describes the change of the population in the phase space due to the flux of particles against the density gradient (second term in the lhs of Eq. \eqref{collint1}) and due to collisions (rhs of Eq. \eqref{collint1}). 

The factor $\chi$ in the gain term of the collision integral is specific for Granular Gases. It appears as a result of the transformation from the velocities $\vec{v}^{\,\prime\prime}_1, \vec{v}^{\,\prime\prime}_2$ of the inverse collision to the velocities $\vec{v}_1, \vec{v}_2$. Frequently it is assumed $\varepsilon=\mbox{const.}$ (we will call this {\em simplified collision model}, see section \ref{sec:Visco}), the factor $\chi$ is also a constant for this model, $\chi=1/\varepsilon^2$.  The case of molecular gases is obtained in the elastic limit $\varepsilon =1$.

The collision integral has an important property: Given 
\begin{equation}
\left< \psi\left(t\right) \right> \equiv \int d\vec{v}_1 \psi \left(\vec{v}\,\right) f\left(\vec{v}, t\right)  
\end{equation}
is the average value of some function $\psi \left(\vec{v}\,\right)$ and $\Delta \psi\left(\vec{v}_i\right) \equiv \left[\psi\left(\vec{v}^{\,\prime}_i\right)-\psi\left(\vec{v}_i\right) \right]$ denotes the change of $\psi\left( \vec{v}_i\right)$ due to a direct collision. Then \cite{ChapmanCowling:1970,FerzigerKaper:1972}
\begin{multline}
\label{deraver} \frac{d}{dt} \left< \psi\left(t\right) \right>
=\int d\vec{v}_1 \psi \left(\vec{v}_1\right)
\frac{\partial}{\partial t} f\left(\vec{v}_1, t\right) = \int
d\vec{v}_1 \psi \left(\vec{v}_1\right) I\left(f,f\right)\\
=\frac{\sigma^{d-1}}{2}\! \int\! d\vec{v}_1d\vec{v}_2 \!\int\!
d\vec{e}\, \Theta\left(-\vec{v}_{12} \cdot \vec{e}\,\right)
\left|\vec{v}_{12} \cdot \vec{e}\,\right| f\left(\vec{v}_1,
t\right)f\left(\vec{v}_2, t\right) \\ \times\Delta \left[ \psi\left(
\vec{v}_1\right)+ \psi\left( \vec{v}_2\right) \right] \, .
\end{multline}
Hence, the time derivative of any avarage quantity  $\psi \left(\vec{v}\right)$ is expressed in terms of a function of the collision integral. The integration is performed over two velocities $\vec{v}_1$, $\vec{v}_2$ and over the unit vector $\vec{e}$. The integrand is the product of two distribution functions and of the expression $\Delta \left[ \psi\left( \vec{v}_1\right)+ \psi\left( \vec{v}_2\right) \right]$, which depends on the above three vectors since the after-collision velocities are given by Eq. \eqref{Vel:15a}.

We will show below that all important properties of dilute homogeneous and non-homogeneous Granular Gases may be formulated in terms of expressions of the collision integral of the same structure as in Eq. \eqref{deraver}. We will call these functions {\em kinetic integrals}. The structures of the kinetic integrals do not depend on the collision model, which allows to elaborate a general approach for gases of elastic particles, for dissipative gases with a simplified collisional model $\varepsilon=\mbox{const.}$, as well as for gases of particles which collide via a realistic collision model as discussed in section \ref{sec:Visco}.   

The aim of this paper is to describe a general approach for the
analytical evaluation of kinetic integrals by means of symbolic
algebra. To illustrate its application we compute various
characteristics of homogeneous and inhomogeneous Granular Gases.

\section{A simple example}
\label{sec:example}

We consider a homogeneous Granular Gas where the velocity distribution function is independent of $\vec{r}$. The granular temperature $T$ is defined by the second moment of the velocity distribution function:
\begin{equation}
\label{deftemp1}
\frac{d}{2} n T\left(t\right)=\int d \vec{v} \frac{m v^2}{2} f\left(\vec{v},t\right)\,, 
\end{equation}
which evolves due to \cite{EsipovPoeschel:1995,NoijeErnst:1998,BrilliantovPoeschelStability:2000,BrilliantovPoeschel:2000visc} 
\begin{equation}
\frac{dT}{dt}= - \zeta T \qquad\qquad {\rm with } \qquad \qquad
 \zeta \equiv \frac{2}{d} n \sigma^{d-1} \sqrt{ \frac{2T}{m}} \mu_2 \, , 
\label{eq:zeta_def_hom1}
\end{equation}
where $\zeta$ is the cooling coefficient of the homogeneous gas, $n$ is the number density, $m$ is the particle mass and the coefficient $\mu_2$ is the second moment of the dimensionless collision integral, with the general definition \cite{NoijeErnst:1998,BrilliantovPoeschelStability:2000,BrilliantovPoeschel:2000visc} 
\begin{equation}
\label{mup1}
\mu_p=-\frac12 \int d\vec{c}_1\int d\vec{c}_2 \int d\vec{e} \,
\Theta\left(-\vec{c}_{12} \cdot \vec{e}\,\right) \left|\vec{c}_{12} \cdot \vec{e}\,\right| 
\tilde{f}\left(c_1\right) \tilde{f}\left(c_2\right) \Delta \left(c_1^{\,p}+c_2^{\,p}\right) \,.
\end{equation}
Here $\tilde{f}\left(c\right)$ is the dimensionless distribution function
\begin{equation}
f(v,t)=\frac{n}{\vec{v}_T^{\,d}}\tilde{f}\left(\frac{v}{\vec{v}_T} \right)\,,  
\end{equation}
which depends on the reduced velocity $\vec{c} \equiv\vec{v} / \vec{v}_T$, with 
\begin{equation}
\label{eq:thermalvel}
\vec{v}_T(t) \equiv \sqrt{\frac{2T(t)}{m}}  
\end{equation}
being the thermal velocity of the gas. 

For Granular Gases the velocity distribution function deviates from the Maxwell distribution. This deviation is characterized by the coefficients of the Sonine polynomial expansion:
\begin{equation}
\label{Soninexp}
\tilde{f}\left(\vec{c}\,\right)=\phi\left(c\right) \left[1 + \sum_{p=1}^{\infty} a_p S_p\left(c^2\right) \right]\,
\end{equation}
where the leading term $\phi\left(c\right) \equiv \pi^{-d/2} \exp\left(-c^2\right) $ is the reduced Maxwell distribution, $a_p$ are numerical coefficients and the Sonine polynomials read for general dimension $d$: 
\begin{equation}
\label{eq:Soninedef}
  \begin{split}
S_0\left(x\right)&=1\\
S_1\left(x\right)&=-x + \frac{d}{2}\\
S_2\left(x\right)&=\frac12 x^2 - \frac12 \left(d+2\right) x + \frac18 \left(d+2\right)d\\
\mbox{etc}.       
  \end{split}
\end{equation}
For small dissipation ($\varepsilon\lesssim 1$) the deviations of the velocity distribution function from the Maxwell distribution are small. Therefore, we approximate
\begin{equation}
\label{fa2x}
\tilde{f}\left(c\right) \simeq \phi \left(c\right) \left[ 1+a_2S_2\left(c^2\right)\right] \, , 
\end{equation} 
where we take into account that $a_1=0$ due to the definition of temperature \cite{NoijeErnst:1998,BrilliantovPoeschelStability:2000,BrilliantovPoeschel:2000visc}. 

If we further neglect terms ${\mathcal O}\left(a_2^2\right)$, the second moment of the dimensionless collision integral reads
\begin{multline}
\label{mupa2x}
\mu_2=\int d\vec{c}_1\int d\vec{c}_2 \int d\vec{e}\, 
\Theta\left(-\vec{c}_{12} \cdot \vec{e}\,\right) \left|\vec{c}_{12} \cdot \vec{e}\,\right| \phi\left(c_1\right) \phi\left(c_2\right) \times \\
\times -\frac12 \left\{1+a_2\left[S_2\left(c_1^2\right)+S_2\left(c_2^2\right) \right] \right\}\Delta \left(c_1^2+c_2^2\right)\, .  
\end{multline}
This expression is of the form
\begin{equation}
\label{mupa2xx}
\mu_2=\int d\vec{c}_1\int d\vec{c}_2 \int d\vec{e}\, 
\Theta\left(-\vec{c}_{12} \cdot \vec{e}\,\right) \left|\vec{c}_{12} \cdot \vec{e}\,\right| \phi\left(c_1\right) \phi\left(c_2\right) \text{Expr(\dots)} 
\end{equation}
with 
$\text{Expr}=-\frac12 \left\{1+a_2\left[S_2\left(c_1^2\right)+S_2\left(c_2^2\right) \right] \right\}\Delta \left(c_1^2+c_2^2\right)$ 

We apply the following Maple program to obtain $\mu_2$ and, consequently, the cooling rate of the Granular Gas due to Eq. \eqref{eq:zeta_def_hom1}:

\noindent\hspace*{0.0cm}\fbox{\begin{minipage}{11.5cm}
{\footnotesize 
\input{AppenDemomu2}
}
\end{minipage}}
\medskip
\medskip

\noindent 
This example is intended to demonstrate the application of symbolic algebra using the package \verb|KineticIntegral.m| for the evaluation of kinetic integrals without explaining the details. The mathematical method as well as the according Maple package are subject of the next sections.

Keeping terms ${\mathcal O}\left(a_2^2\right)$ we obtain the result
\begin{equation}
\label{mu2A}
\mu_2=\sqrt{2 \pi} \left(1-\varepsilon^2\right)\left(1+\frac{3}{32}a_2 \right)^2 \, ,  
\end{equation}
which yields with Eq. \eqref{eq:zeta_def_hom1} the temperature evolution law (Haff's law) 
\begin{equation}
\label{T(t)forconst}
T\left(t\right)=\frac{T_0}{\left( 1+t/\tau_0 \right)^2}\,  
\end{equation}
with the characteristic time $\tau_0$ 
\begin{equation}
\label{tau0forconst}
\tau_0^{-1}=\frac{\mu_2}{3 \sqrt{8 \pi}} \tau_c\left(0\right)^{-1}=
\frac16 \left(1-\varepsilon^2\right) \left(1+\frac{3}{16}a_2+\frac{9}{1024}a_2^2 \right)\tau_c\left(0\right)^{-1} \, ,  
\end{equation}
where the mean collision time is given by 
\begin{equation}
  \label{eq:meancolltimet}
\tau_c\left(t\right)^{-1} \equiv 4 \sqrt{\pi}\sigma^2 n\sqrt{\frac{T(t)}{m}} \, .
\end{equation}

This example shows that with few lines only we can derive the
temperature decay of a Granular Gas. Since the first part of the
program is identical in all applications, only the main part is
given below in the text. The program above then reads

\newprog
{\scriptsize 
\begin{Verbatim}
DefDimension(3);           
DefJ();
Expr:=Delta2*(-1/2**(1+a2*(S(2,c1p)+S(2,c2p)))):
mu2:=getJexpr(0,Expr,0);
\end{Verbatim}
}

\section{Granular Gases of viscoelastic particles}
\label{sec:Visco}

It has been shown experimentally, e.g. \cite{BridgesHatzesLin:1984}, as well as theoretically \cite{Ramirez:1999} that the coefficient of restitution of realistic particles is not a material constant but it depends on the impact velocity. For the case of viscoelastic interaction the coefficient of restitution can be derived from the detailed analysis of the interaction of colliding spheres. The force between viscoelastic spheres of radii $R_i$ and $R_j$ which deform by $\xi\equiv \left|\vec{r}_i - \vec{r}_j\right|-R_i-R_j$ at the rate $\dot{\xi}$ is given as a superposition of Hertz's contact force \cite{Hertz:1882} and a viscous force \cite{BrilliantovSpahnHertzschPoeschel:1994}:
\begin{equation}
  \label{eq:ViscoForce}
   F=\frac{2\,Y\,\sqrt{R^{\rm eff}}}
{3\left(1-\nu^2\right)}\left(\xi^{3/2} +  \frac32A\sqrt{\xi}\dot{\xi}\right)\equiv \alpha\xi^{3/2}+\frac32A  \alpha \sqrt{\xi}\dot{\xi}\,, 
\end{equation}
where $Y$, $\nu$ are respectively the Young modulus and the Poisson ratio and $R^{\rm eff}\equiv R_iR_j/\left(R_i+R_j\right)$. The dissipative constant $A$ is a function of the bulk viscosities of the particles material (see \cite{BrilliantovSpahnHertzschPoeschel:1994} for details). Integrating Newton's equation of motion for a particle collision the coefficient of restitution is obtained as a function of the normal component of the impact velocity $g\equiv\left|\vec{e} \cdot \left(\vec{v}_1-\vec{v}_2\right)\right|$ \cite{SchwagerPoeschel:1998,Ramirez:1999}
\begin{equation}
\varepsilon=1- C_1 A \alpha^{2/5} g^{\,1/5} +\frac35 C_1^2 A^2  \alpha^{4/5} g^{\,2/5}  \mp \dots
\label{epsinv}
\end{equation}
The numerical coefficient $C_1$ reads
\begin{equation}
  C_1=\frac{ \sqrt{\pi}}{2^{1/5}5^{2/5}} \, \frac{\Gamma \left(3/5\right)}{\Gamma\left(21/10\right)}\,.
\label{eq:defC1}
\end{equation}

Using the definitions of the thermal velocity, Eq. \eqref{eq:thermalvel}, and the reduced velocity $\vec{c} \equiv\vec{v} / \vec{v}_T$ the impact-velocity dependent coefficient of restitution, Eq. \eqref{epsinv}, may be written in terms of the reduced velocity:
\begin{equation}
\label{epsc12}
\varepsilon= 1-C_1\delta^{\prime}\left(t\right) \left|\vec{c}_{12} \cdot \vec{e} \,\right|^{1/5}+
\frac{3}{5}C_1^2 \delta^{\prime \, 2}\left(t\right) \left|\vec{c}_{12} \cdot 
\vec{e}\, \right|^{2/5}\, ,
\end{equation} 
where 
\begin{equation}
\label{deltaprime}
\delta^{\, \prime} (t) \equiv A \alpha^{2/5} \left[2T(t)\right]^{1/10} \equiv \delta \left(\frac{2T(t)}{T_0} \right)^{1/10} \, , 
\end{equation}
with $\delta \equiv A \alpha^{2/5}T_0^{1/10}$ and with $T_0$ being the initial temperature. We will need this form below.

The derivation of the kinetic properties of Granular Gases of viscoelastic particles is significantly more complicated than for the case of the simplified coefficient of restitution ($\varepsilon=\text{const.}$), see e.g. \cite{BrilliantovPoeschel:2000visc,BrilliantovPoeschel:2001roy}. The main reason of these complications is the factor $\chi$ in Eq. \eqref{collint1} which is not a constant anymore but a complicated function of the impact velocity \cite{BrilliantovPoeschel:2000visc}. The elaboration of the kinetics of Granular Gases of viscoelastic particles becomes very technical and was the main motivation for the application of computational symbolic algebra as presented here.

In this paper we consider systems of three-dimensional spheres, i.e., for the case of viscoelastic spheres the particles interact by the force \eqref{eq:ViscoForce}. Hence, for $d=2$ we describe a gas of three-dimensional  viscoelastic spheres whose motion is restricted to a plane. The coefficient of restitution for $d=2$, i.e. for viscoelastic disks, has been derived recently \cite{Schwager2001,Schwager2003}.

\section{Evaluation of kinetic integrals}
\subsection{Definition of the basic integrals}

The evaluation of kinetic integrals requires the computation of a sum of simpler integrals which we call {\em basic integrals}. These basic integrals may be computed analytically. Kinetic integrals are then decomposed into basic integrals. Although this decomposition is straightforward, the number of basic integrals may be as large as several thousands.  Thus in practice it is not feasible to evaluate kinetic integrals manually.  Let us consider a simple kinetic integral and show how it can be expressed in terms of basic integrals. In what follows we elaborate the computation of kinetic integrals for general dimension $d$. Below the application of the method will
be illustrated only for the most important case $d=3$. The corresponding results for $d=2$ can be obtained by 
changing the command \verb|DefDimension(3)| into \verb|DefDimension(2)| in the beginning of the programs.

We assume that the dissipation is small enough to keep only first nonvanishing term in the Sonine polynomials expansion Eq. \eqref{Soninexp}, i.e. we assume that the approximation
\begin{equation}
\label{fa2}
\tilde{f}\left(c\right) \simeq \phi \left(c\right) \left[ 1+a_2S_2\left(c^2\right)\right] \, .
\end{equation} 
is adequate. With this approximation the moments of the dimensionless collision integral $\mu_p$ reads
\begin{multline}
\label{mupa2}
\mu_p=-\frac12 \int d\vec{c}_1\int d\vec{c}_2 \int d\vec{e}\, 
\Theta\left(-\vec{c}_{12} \cdot \vec{e}\,\right) \left|\vec{c}_{12} \cdot \vec{e}\,\right| \phi\left(c_1\right) \phi\left(c_2\right) \times \\[0.2cm]
\times\left\{1+a_2\left[S_2\left(c_1^2\right)+S_2\left(c_2^2\right) \right] + a_2^2\,S_2\left(c_1^2\right)S_2\left(c_2^2\right) \right\}\Delta \left(c_1^{\,p}+c_2^{\,p}\right)\, .  
\end{multline}
To evaluate the kinetic integral in Eq. \eqref{mupa2}  it is convenient to use the center of mass velocity $\vec{C}$ and the relative velocity $\vec{c}_{12}$:
\begin{equation}
\label{Cc12viac1c2}
\vec{c}_{1}=\vec{C}+\frac12 \vec{c}_{12}\,, \qquad \qquad \vec{c}_{2}=\vec{C}-\frac12 \vec{c}_{12}\,.
\end{equation}
It is easy to check that the Jacobian of the transformation Eq. \eqref{Cc12viac1c2} equals unity. The product of the Maxwell distributions $\phi\left(\vec{c}_{1}\right) \phi\left(\vec{c}_{2}\right)$ transforms into a product of a Maxwell distributions for the center of mass (an effective particle of double mass) and of a Maxwell distribution for the relative motion (an effective particle of half mass)
\begin{equation}
\label{MaxtoMax}
\begin{split}
\phi\left(\vec{c}_{1}\right) \phi\left(\vec{c}_{2}\right) \to&
\frac{1}{\left(2 \pi\right)^{d/2}} \exp \left(-\frac12 c_{12}^2
\right) \left(\frac{2}{\pi} \right)^{d/2} \exp \left( -2C^2
\right)\\ & \equiv \phi\left(\vec{c}_{12}\right)
\phi\left(\vec{C}\right)\,.
 \end{split}
\end{equation}
The terms in squared brackets in Eq. \eqref{mupa2} hence transform into
\begin{multline}
\label{S2S2Cc12}
\left[ S_2\left(c_1^2\right)+S_2\left(c_2^2\right) \right]=\\
C^4+\left(\vec{C} \cdot \vec{c}_{12} \right)^2 +\frac{1}{16}c_{12}^4
+\frac12  C^2 c_{12}^2-5 C^2-\frac54 c_{12}^2 +\frac{15}{4}\, .
\end{multline}
To find the quantities $\Delta\left(c_1^p+c_2^p\right)$ ($p=2,4$) we rewrite the collision law Eq. \eqref{Vel:15a} in terms of the $\vec{C}$ and $\vec{c}_{12}$ too:
\begin{equation}
  \begin{split}
\vec{c}^{\,\prime}_1=&\vec{C}+\frac12 \vec{c}_{12}-\frac12 \left(1+\varepsilon\right) \left(\vec{c}_{12} \cdot \vec{e}\,\right)\vec{e}\\
\vec{c}^{\,\prime}_2=&\vec{C}-\frac12 \vec{c}_{12}+\frac12 \left(1+\varepsilon\right) \left(\vec{c}_{12} \cdot \vec{e}\,\right)\vec{e} \, .  
  \end{split}
\label{directcollviaCc}
\end{equation}
Then straightforward algebra yields
\begin{equation}
\label{Deltac1c2p2}
\Delta\left(c_1^2+c_2^2\right)=
-\frac12 \left(1-\varepsilon^2\right) \left(\vec{c}_{12} \cdot \vec{e}\, \right)^2 
\end{equation}
and
\begin{equation}
\label{Deltac1c24}
\begin{split}
\Delta\left(c_1^4+c_2^4\right)=& 2\left(1+\varepsilon\right)^2\left(\vec{c}_{12} \cdot \vec{e}\, \right)^2
\left(\vec{C} \cdot \vec{e}\, \right)^2 +\frac18 \left(1-\varepsilon^2\right)^2 \left(\vec{c}_{12} \cdot \vec{e}\, \right)^4\\
 &-\frac14 \left(1-\varepsilon^2\right) \left(\vec{c}_{12} \cdot \vec{e}\, \right)^2 \vec{c}_{12}^{\,\,2} - \left(1-\varepsilon^2\right) \vec{C}^{\,2}  \left(\vec{c}_{12} \cdot \vec{e}\, \right)^2 \\
&- 4 \left(1+\varepsilon\right) \left(\vec{C} \cdot \vec{c}_{12} \right) \left(\vec{C} \cdot \vec{e}\, \right) 
\left(\vec{c}_{12} \cdot \vec{e}\, \right)\,.
\end{split}
\end{equation}
Combining the terms in Eqs. \eqref{S2S2Cc12}, \eqref{Deltac1c2p2}, and \eqref{Deltac1c24} the integrands for $\mu_2$ and $\mu_4$ in Eq. \eqref{mupa2} contain products of the Maxwell distributions for $C$ and $c_{12}$ and the factors $C^{\,k}$, $c_{12}^{\,l}$, $\left(\vec{c}\cdot \vec{c}_{12}\right)^m$ with exponents $k$, $l$, $m$, and similar other factors. Hence we define the  {\em basic integral } by 
\begin{multline}
\label{basicint}
J_{k,l,m,n,p,\alpha}\equiv \int d \vec{c}_{12} \int d \vec{C} \int d\vec{e}\, \Theta\left(-\vec{c}_{12} \cdot \vec{e}\,\right) \left|\vec{c}_{12} \cdot \vec{e}\,\right|^{1+\alpha}\\[0.2cm] 
\times \, \phi\left(c_{12}\right) \phi\left(C\right) C^k c_{12}^l \left(\vec{C} \cdot \vec{c}_{12} \right)^m \left( \vec{C} \cdot \vec{e}\, \right)^n \left(\vec{c}_{12} \cdot \vec{e}\, \right)^p \, , 
\end{multline}
and express $\mu_p$ via these quantities. In particular $\mu_2$ reads in linear approximation with respect to $a_2$:
\begin{multline}
\label{mu2viabasint}
\mu_2=\frac14 \left(1-\varepsilon^2\right)\left[ J_{0,0,0,0,2,0} +
a_2 \left( J_{4,0,0,0,2,0} + J_{0,0,2,0,2,0} + \frac{1}{16}J_{0,4,0,0,2,0} \right.\right.\\\left.\left.
+\frac12 J_{2,2,0,0,2,0} - 5 J_{2,0,0,0,2,0} -\frac54 J_{0,2,0,0,2,0} 
+\frac{15}{4} J_{0,0,0,0,2,0} \right) \right]\,.
\end{multline}
The corresponding expression for $\mu_4$ contains already 35 basic integrals. Keeping the second order terms with respect to $a_2$ yields expressions which contain 60 basic integrals for $\mu_2$ and $328$ basic integrals for $\mu_4$. To handle such expressions one should obviously use computer algebra.

\subsection{Computation of the basic integrals}

\subsubsection{Solution of the angular integrals}
The solution of the basic integral as defined by Eq. \eqref{basicint} is straightforward for $n=0$, however, for $n=1,2$ it requires some tricks (see, e.g., \cite{resibua}). The presented approach may be generalised to $n > 2$, here we need only $n=0,1,2$. 

In the $d$-dimensional space the Cartesian coordinates of a vector $\vec{r}=\left(x_1,\dots,x_d\right)$ are related with the spherical coordinates, expressed by the absolute value of the vector $r \ge 0 $ and $d-1$ angles $\theta_1, \, \theta_2, \, \ldots \theta_{d-1}$ by the transformation
\begin{equation}
  \begin{split}
    x_1  & =r\sin \theta_1 \sin \theta_2 \ldots  \sin \theta_{d-1} \\ 
    x_m  & =r \cos \theta_{m-1} \prod_{k=m}^{d-1} \sin \theta_k\,, \qquad \qquad m =2,3, \ldots d-1 \\
    x_{d}& =r \cos \theta_{d-1} \,,  
  \end{split}
\label{eq:sphcoor_App}
\end{equation}
where 
\begin{equation}
  \begin{split}
    &0 \le \theta_1  \le 2 \pi \\
    &0 \le \theta_m  \le \pi\,,  \qquad \qquad m =2,3, \ldots d-1  \, . 
  \end{split}
\label{eq:sphcoor2_App}
\end{equation}
Hence, the infinitesimal solid angle reads
\begin{equation}
  d \vec{e} = \prod_{k=1}^{d-1} \sin^{k-1} \theta_k d \theta_k  \, .  
\label{eq:sphcoor3_App}
\end{equation}
Correspondingly we obtain for the surface area of a $d$-dimensional unit sphere
\begin{equation}
 \begin{split}
  \Omega_d & =\int d \vec{e} = \int_0^{2\pi} d\theta_1 \int_{0}^{\pi} \sin \theta_2 d\theta_2  \int_{0}^{\pi} \sin^2 \theta_3 d\theta_3 
            \ldots \int_{0}^{\pi} \sin^{d-2}  \theta_{d-1} d\theta_{d-1} \\[0.2cm]
           & =2 \pi \, \, \frac{\Gamma \left( 1 \right) \Gamma \left( \frac12 \right)}{\Gamma\left(\frac32\right)} \, \, 
              \frac{\Gamma \left( \frac32 \right) \Gamma \left( \frac12 \right)}{\Gamma\left( 2 \right)} \, \, 
              \frac{\Gamma \left( 2  \right) \Gamma \left( \frac12 \right)}{\Gamma\left( \frac52 \right)} \, \,
              \ldots \frac{\Gamma\left(\frac{d-1}{2}\right) \Gamma\left(\frac12\right)}{\Gamma\left(\frac{d}{2}\right)} \\[0.2cm]
           & = 2 \pi \left[ \Gamma \left( \frac12 \right) \right]^{d-2} \frac{1}{\Gamma\left(\frac{d}{2}\right)}
             =  \frac{2 \pi^{d/2}}{\Gamma \left(\frac{d}{2}\right)} \,,
 \end{split}         
\label{eq:Omegad_App}
\end{equation}
where we take into account that $\Gamma \left( 1 \right) =1 $, $\Gamma \left(1/2 \right) = \sqrt{\pi}$ and we use the trigonometric integrals
\begin{equation}
\label{eq:int2_App}
\int_{0}^{\pi/2}  \sin^l \theta \, \cos^n \theta  d\theta  = 
\frac12 \, 
\frac{\Gamma\left( \frac{l+1}{2} \right) \Gamma \left(\frac{n+1}{2}\right)}{\Gamma\left(\frac{l+n}{2}+1 \right)} \, .
\end{equation}
and
\begin{equation}
\int_{0}^{\pi} \sin^l \theta d\theta = \frac{\Gamma\left( \frac{l+1}{2} \right) \Gamma \left(\frac{1}{2}\right)}{\Gamma\left(\frac{l}{2}+1 \right)} \, . 
\label{eq:int_sin}
\end{equation}

Before calculating  the Basic Integral we evaluate some more simple expressions. We will need the angular integral 
\begin{equation}
  \beta_m  =  \int d \vec{e} \, \Theta \left( \vec{e} \cdot \hat{\vec{g}} \right) \left( \vec{e} \cdot \hat{\vec{g}} \right)^m   
           = (-1)^m  \int d \vec{e} \, \Theta \left( -\vec{e}  \cdot \hat{\vec{g} }  \right) \left( \vec{e} \cdot \hat{\vec{g}} \right)^m  \, ,  
\label{eq:defbetn_App}
\end{equation}
where  vectors with a hat denote unit vectors (e.g. $\hat{\vec{g} } \equiv \vec{g} /g$). In Eq. \eqref{eq:defbetn_App} we change the variables $\vec{e} \to - \vec{e}$ to get the expression for $\beta_m$ which will be mainly used in the calculations. Let us choose the coordinate axis $OX_d$ along $\vec{g}$, so that $\vec{e} \cdot \hat{\vec{g}} =\cos \theta_{d-1}$.  Using Eqs. \eqref{eq:int_sin} \eqref{eq:int2_App} the above properties of the Gamma function and Eq. \eqref{eq:sphcoor3_App} for the infinitesimal solid angle $d \vec{e}$  we obtain
\begin{equation}
 \begin{split}
   \beta_m & = \int d \vec{e} \, \Theta \left( \vec{e} \cdot \hat{\vec{g}} \right) \left( \vec{e} \cdot \hat{\vec{g}} \right)^m \\
           & = \int\limits_0^{2\pi} d\theta_1 \int\limits_{0}^{\pi} \sin \theta_2 d\theta_2  \int\limits_{0}^{\pi} \sin^2 \theta_3 d\theta_3 
            \ldots \int\limits_{0}^{\pi/2} \sin^{d-2}  \theta_{d-1} \cos^m \theta_{d-1}d\theta_{d-1} \\
           & =2 \pi \, \, \frac{\Gamma \left( 1 \right) \Gamma \left( \frac12 \right)}{\Gamma\left(\frac32\right)} \, \, 
               \frac{\Gamma \left( \frac32 \right) \Gamma \left( \frac12 \right)}{\Gamma\left( 2 \right)} \, \, 
               \ldots \frac12 \, \frac{\Gamma\left(\frac{d-1}{2}\right) \Gamma\left(\frac{m+1}{2}\right)}{\Gamma\left(\frac{d+m}{2}\right)} \\
           &=  \pi^{(d-1)/2} \frac{ \Gamma \left( \frac{m+1}{2} \right) }{ \Gamma \left( \frac{m+d}{2} \right)} \, ,  
\end{split}
\label{eq:betn1_App}
\end{equation} 
where we take into account that the integration over $\theta_{d-1}$ is to be performed in the interval $[0, \, \pi/2]$ due  to the factor $\Theta \left( \vec{e} \cdot \hat{\vec{g}} \right)$. Note that $\beta_m$ does not depend on the vector $\hat{\vec{g} }$.  
Similarly we obtain 
\begin{equation}
  \begin{split}
  \int d \vec{e} \left(\vec{e} \cdot \hat{\vec{g}} \right)^m & = \int d \vec{e} \left[  \Theta \left( \vec{e} \cdot \hat{\vec{g}} \right) 
                                        + \Theta \left( -\vec{e}  \cdot \hat{\vec{g} }  \right) \right] \left( \vec{e} \cdot \hat{\vec{g}} \right)^m  
                                     = \left[ 1 + (-1)^m \right] \beta_m  \\
                                     & = \int d \hat{\vec{g}}  \left( \hat{\vec{g}}  \cdot \hat{\vec{C}}  \right)^m  \, ,  
\end{split}
\label{eq:alpham_App}
\end{equation}
where we take into account that $\vec{e}$ is the integration variable and that $\beta_m$ is independent of $\hat{\vec{g} }$. This allows to change the notation for the integration variable and for $\hat{\vec{g} }$: $\vec{e} \to \hat{\vec{g} }$ and $\hat{\vec{g} } \to \hat{\vec{C} }$. 

\subsubsection{Reduction of the basic integrals}

We also need the integrals 
\begin{multline}
\label{eq:gammar_App}
   \int d \vec{C} C^{\,r} \phi(C) = \int_0^{\infty} c^{\,d-1+r} \left( \frac{2}{\pi} \right)^{d/2} e^{-2\,C^{\,2}} \, dC  \, \int d \hat{\vec{C}} \\
                         = \left( \frac{2}{\pi} \right)^{d/2} \frac14 \frac{\sqrt{2} \Gamma \left( \frac{d+r}{2} \right) }{2^{\frac{d+r-1}{2}}} \, \Omega_d = 2^{-r/2} \frac{ \Gamma \left( \frac{d+r}{2} \right) }{\Gamma \left( \frac{d}{2} \right)} \equiv \gamma_r   
\end{multline}
and 
\begin{multline}
   \int d \vec{g} g^r \phi(g)\\ = \int_0^{\infty} g^{d-1+r} \left( \frac{1}{2\pi} \right)^{d/2} e^{-g^2/2} \, dg  \, \int d \hat{\vec{g}}
                         = 2^{\,r/2} \frac{ \Gamma \left( \frac{d+r}{2} \right) }{\Gamma \left( \frac{d}{2} \right)} = 2^{\,r} \gamma_r\,, 
\label{eq:deltar_App}
\end{multline}
where again we have used $\int d \hat{\vec{g}} = \int d \hat{\vec{C}} = \int d \vec{e} = \Omega_d$ according to Eq. \eqref{eq:Omegad_App}.

For $n=0$ the definition of the basic integral, Eq. \eqref{basicint},  reads (with $\vec{g} \equiv \vec{c}_{12}$)
\begin{equation}
\label{Jn0eval}
J_{k,l,m,0,p,\alpha}= \int d \vec{g}  \int d \vec{C} \phi(g) \phi(C) C^k g^l (\vec{C} \cdot \vec{g}  )^m K_{\alpha,p} (g)\,,
\end{equation}
where we introduce the integral
\begin{equation}
\label{K(g)}
K_{\alpha,p} (g) \equiv \int d \vec{\mu} \,  \left|\vec{e} \cdot \vec{g} \,\right|^{\alpha} \left( \vec{e} \cdot \vec{g}\, \right)^p \,,
\end{equation} 
with the short-hand notation $ d \vec{\mu} \equiv d \vec{e} \, \Theta \left( -\vec{e} \cdot \vec{g}\, \right) \left|\vec{e} \cdot \vec{g}\, \right|$. Similarly, for $n=1$ one can write
\begin{equation}
\label{Jn1eval}
J_{k,l,m,1,p,\alpha}= \int d \vec{g}  \int d \vec{C} \phi(g) \phi(C) 
C^k g^l \left( \vec{C} \cdot \vec{g}\,  \right)^m \left( \vec{C} \cdot \vec{I}_{\alpha,p} (g) \right) \,,
\end{equation}
with the vectorial integral
\begin{equation}
\label{I(g)}
\vec{I}_{\alpha,p}(g) \equiv \int d \vec{\mu} \, \vec{e} \,\left|\vec{e} \cdot \vec{g}\, \right|^{\alpha} \left( \vec{e} \cdot \vec{g}\, \right)^p \, . 
\end{equation} 
Finally,  for $n=2$:
\begin{equation}
\label{Jn2eval1}
J_{k,l,m,2,p,\alpha}= \int d \vec{g}  \int d \vec{C} \phi(g) \phi(C) 
C^k g^{\,l} \left( \vec{C} \cdot \vec{g}\,  \right)^m \, \vec{C} \cdot \hat{H}_{\alpha,p}(g) \cdot \vec{C}
\end{equation}
where the dyad  $\hat{H}_{\alpha,p}(g)$ is given by
\begin{equation}
\label{H(g)}
\hat{H}_{\alpha,p}(g) \equiv \int d \vec{\mu} \,
\vec{e} \, \vec{e} \left|\vec{e} \cdot \vec{g}\, \right|^{\alpha}
\left( \vec{e} \cdot \vec{g} \,\right)^p
\end{equation}
and where we take into account that according to the vector-dyadic product
\begin{equation}
\vec{C} \cdot \vec{e}\, \vec{e} \cdot \vec{C} = \left( \vec{C} \cdot \vec{e} \right) \left( \vec{e} \cdot \vec{C} \right) = \left( \vec{C} \cdot \vec{e} \right)^2\,.   
\end{equation}
With the quantities introduced above we can express the basic integral step by step. 

\subsubsection{Solution of the auxiliary vectorial and tensorial integrals}

The integral $K_{\alpha,\,p} (g)$ reads then
\begin{equation}
\label{K(g)1}
 \begin{split}
  K_{\alpha,p} (g) & = \int d \vec{e} \, \Theta \left( -\vec{e} \cdot \vec{g}\, \right) \left| \vec{e} \cdot \vec{g}\, \right|^{\alpha+1} \left( \vec{e} \cdot \vec{g}\, \right)^p \\
                   & = (-1)^p g^{\,p+\alpha+1} \int d \vec{e}  \, \Theta \left( \vec{e} \cdot \hat{\vec{g}}  \right)  \left( \vec{e} \cdot \hat{\vec{g}} \right)^{p+\alpha+1} \\
                   & = (-1)^p g^{\,p+\alpha+1} \beta_{\,p+\alpha+1} \,  
 \end{split}
\end{equation} 
with the definition Eq. \eqref{eq:defbetn_App} of $\beta_m$. Consider now the vectorial integral $\vec{I}_{\alpha,p}(g) = \vec{g} G(g)$. The result of the integration over $\vec{e}$ must be a vector. Due to the symmetry of the problem this vector should be directed along the vector $\vec{g}$ since there is no other preferred  direction. Thus we write
\begin{equation}
\vec{I}_{\alpha,p}(g) = \vec{g}\, G(g)\,.
\end{equation}
The function $G(g)$ may be found by multiplying both sides of Eq. \eqref{I(g)} by $\vec{g}$:
\begin{multline}
\label{I(g)1} 
 \vec{g} \cdot \vec{I}_{\alpha,p}(g) = g^{\,2} G(g)= \int d \vec{e} \, \Theta \left( -\vec{e} \cdot \vec{g}\, \right) \left| \vec{e} \cdot \vec{g}\, \right|^{\alpha+1} \left( \vec{e} \cdot \vec{g}\, \right)^{p+1} = K_{\alpha,p+1} (g)\,.
\end{multline}
Therefore, $G(g)=g^{-2}K_{\alpha,p+1} (g)$ and thus with Eq. \eqref{K(g)1} for $K_{\alpha,p} (g)$ we obtain
\begin{equation}
\label{I(g)2}
   \vec{I}_{\alpha,p}(g) = (-1)^{p+1} \beta_{\,p+\alpha+2} \, g^{\,p+\alpha} \vec{g} \,.
\end{equation}
The integrand in Eq. \eqref{H(g)} is a dyad  proportional to the dyad $\vec{e}\, \vec{e}$. Since integration over $\vec{e}$ does not change the symmetry, we expect that the components of the resultant dyad  $\hat{H}_{\alpha,p}(g)$ are composed of the components of the vector $\vec{g}$, the only vectorial quantity which is contained. Since the result may also have a non-vectorial part, the following form of $\hat{H}_{\alpha,p}(g)$ is suggested: 
\begin{equation}
\label{H(g)1}
\hat{H}(g)=A(g)\, \vec{g}\, \vec{g} + B(g)g^{\,2}\hat{U}\,,
\end{equation}
where $\hat{U}$ is a unit dyad  (i.e. the unit diagonal matrix) and the functions  $A(g)$ and $B(g)$ are to be determined. Multiplying $\hat{H}$ scalarly from both sides by $\vec{g} $ we obtain the first equation for $A(g)$ and $B(g)$:
\begin{multline}
  \vec{g} \cdot \hat{H} \cdot \vec{g} = A(g)g^4 +B(g)g^4 \\
                             = \int d \vec{e} \, \Theta \left( -\vec{e} \cdot \vec{g}\, \right) \left| \vec{e} \cdot \vec{g}\,\right|^{\alpha+1} \left( \vec{e} \cdot \vec{g}\, \right)^{p+2} 
                             = K_{\alpha,p+2} (g) \, ,  
\label{H(g)2}
\end{multline}
where we use 
\begin{equation}
\vec{g} \cdot \vec{g}\, \vec{g} \cdot \vec{g} = \left(g^2\right)^2 = g^4\qquad \mbox{and} \qquad \vec{g} \cdot \hat{U} \cdot \vec{g} = \vec{g} \cdot \vec{g} =g^2\,.\hfill  
\end{equation}
Then taking the trace of  $\hat{H}$ (i.e. computing the sum of diagonal elements) we arrive at the second equation:
\begin{multline}
  {\rm Tr}~ \hat{H}   = A(g)g^2 + d\, B(g)g^2 
                               = \int d \vec{e} \, \Theta \left( -\vec{e} \cdot \vec{g}\, \right) \left| \vec{e} \cdot \vec{g} \,\right|^{\alpha+1} \left( \vec{e} \cdot \vec{g}\, \right)^p 
                                = K_{\alpha,p} (g) \, . 
\label{H(g)3}
\end{multline}
In Eq. \eqref{H(g)3} (with the summation convention) we take into account
\begin{equation}
{\rm Tr}~ \vec{g}\, \vec{g} = g_ig_i=g^2\,,\qquad 
{\rm Tr}~ \hat{U} = \delta_{ii} = d\,, \qquad
{\rm Tr}~ \vec{e}\, \vec{e} = e_ie_i=e^2 =1\,.
\end{equation}
Solving the set of equations \eqref{H(g)2}, \eqref{H(g)3} for the functions $A(g)$ and $B(g)$ and using Eq. \eqref{K(g)1} for $K_{\alpha,p} (g)$ we obtain
\begin{equation}
\label{H(g)4}
\hat{H} = \frac{(-1)^p\, g^{p+\alpha-1} }{(d-1)}  \left[ \left( d \beta_{\,p+\alpha+3} - \beta_{\,p+\alpha+1} \right) \vec{g}\, \vec{g} +  
                                                 \left( \beta_{\,p+\alpha+1} - \beta_{\,p+\alpha+3} \right) g^2 \hat{U} \right]\,. 
\end{equation}

\subsubsection{Solution for the basic integrals}

With the results  for the scalar integral $K_{\alpha,p} (g)$, for the vectorial integral $\vec{I}_{\alpha,p}(g)$ and for the tensorial integral $\hat{H}_{\alpha,p}(g)$ we can evaluate the basic integrals. For the first basic integral with $n=0$ we find with Eq. \eqref{K(g)1} 
\begin{multline}
\label{eq:Jn0eval2}
J_{k,l,m,0,p,\alpha} = (-1)^p \beta_{p+\alpha+1} \int d \vec{g}  \int d \vec{C} \phi(g) \phi(C) 
                         C^{k+m}  g^{\,l+m+p+\alpha+1}\left(\hat{\vec{C}} \cdot \hat{\vec{g}}  \right)^m \\[0.2cm]
= (-1)^p \beta_{\,p+\alpha+1}\! \int\! d \vec{C} C^{k+m} \phi(C)\! \int\! d\vec{g} g^{l+m+p+\alpha+1} \phi(g)
                      \!\int\! d\hat{\vec{g}} \left( \hat{\vec{g}} \cdot \hat{\vec{C}} \right)^m\!\! \Omega_d^{-1} \, ,
\end{multline}
where we write the vectors $\vec{g}$ and $\vec{C}$ in the form $\vec{g}= g \hat{\vec{g}}$ and  $\vec{C} = C \hat{\vec{C}}$ and for the integration over $d \vec{g}$ use
\begin{multline}\
\label{eq:Jn0eval3}
  \int d\vec{g} F_1(g)F_2\left(\hat{\vec{g}}\,\right) = \int_0^{\infty} g^{d-1} F_1(g) dg \, \int d\hat{\vec{g}} F_2\left(\hat{\vec{g}\,}\right) \\
                               = \int_0^{\infty} g^{d-1}F_1(g) dg \,  
                                   \int d\hat{\vec{g}} \, \, \frac{\int d\hat{\vec{g}} F_2\left(\hat{\vec{g}\,}\right)}{\int  d\hat{\vec{g}}} 
                               = \int d\vec{g} F_1(g) \,  \, \int d\hat{\vec{g}} F_2\left(\hat{\vec{g}\,}\right)  \, \Omega_d^{-1}  
\end{multline}
(recall that $\int d\hat{\vec{g}} = \Omega_d$). The integral over $d \hat{\vec{g}} $ in  Eq. \eqref{eq:Jn0eval2} is $\left[ 1 + (-1)^m \right] \beta_m $, according to Eq. \eqref{eq:alpham_App}, while the integrals over $d \vec{C}$ and over $d \vec{g}$ are given  by Eqs. \eqref{eq:gammar_App} and \eqref{eq:deltar_App}, respectively. As the result we obtain the basic integral for $n=0$:
\begin{multline}
\label{eq:Jn0eval4} 
J_{k,l,m,0,p,\alpha}\\ = (-1)^p \left[ 1 + (-1)^m \right] 2^{\,l+m+p+\alpha+1} \Omega_d^{-1}\, \beta_{\,p+\alpha+1} \beta_{\,m} \gamma_{k+m} \gamma_{l+m+p+\alpha+1} \, . 
\end{multline}
The evaluation of the basic integral for $n=1$ and $n=2$ may be performed by substituting Eqs. \eqref{I(g)2} and \eqref{H(g)4} correspondingly into Eqs. \eqref{Jn1eval} and \eqref{Jn2eval1}. The integrals which are obtained by this substitute are  precisely of the same structure as in Eq. \eqref{eq:Jn0eval2}. Hence, the calculations for $n=1$ and $n=2$ are analogous to the case $n=0$. The final result reads for $n=1$:
\begin{multline}
\label{eq:Jn1eval2}
 J_{k,l,m,1,p,\alpha}  = (-1)^{p+1} \left[ 1 + (-1)^{m+1} \right] 2^{\,l+m+p+\alpha+1} \Omega_d^{-1}  \\ 
\times \,\beta_{\,p+\alpha+2}\, \beta_{\,m+1} \, \gamma_{k+m+1} \gamma_{l+m+p+\alpha+1} 
\end{multline}
and for $n=2$:
\begin{multline}
\label{eq:Jn2eval2}
 J_{k,l,m,2,p,\alpha}  = (-1)^{p} \left[ 1 + (-1)^{m} \right] 2^{\,l+m+p+\alpha+1} \left[ (d-1) \Omega_d \right]^{-1} \\[0.2cm]
\times  \gamma_{k+m+2} \, \gamma_{l+m+p+\alpha+1} \, \left[  \left(d \beta_{p+\alpha+3} -\beta_{p+\alpha+1} \right)  \beta_{m+2}  \right. \\[0.2cm]
\left. + \left( \beta_{p+\alpha+1} - \beta_{p+\alpha+3} \right) \beta_{m} \right] \, . 
\end{multline}
The constants $\beta_m$, $\gamma_m$ and $\Omega_d$ which are contained in the basic integrals have been defined above. 

We gave the solution of the basic integrals for $n \le 2$ with arbitrary other indexes. As illustrated below, these basic integrals are mainly used in the kinetic gas theory. The present approach, however, may be straightforwardly extended. For example, the integral for $n=3$ may be written (similar to \eqref{H(g)} in terms of the tensor $H_{\alpha,p}^{(i,j,k)}(g)=\int
d\mu e_i e_j e_k \left| \vec{e} \cdot \vec{g} \right|^{\alpha}
\left( \vec{e} \cdot \vec{g} \right)^{p}$. Then this tensor and
finally the basic integral may be solved using the same
symmetry-based approach.
\vfill\newpage

\section{Computational formula manipulation to evaluate kinetic integrals}

Kinetic integrals have the structure  
\begin{equation}
\label{eq:App:intMaple}
\int d\vec{c}_1\int d\vec{c}_2 \int d\vec{e}\, 
\Theta\left(-\vec{c}_{12} \cdot \vec{e}\,\right) \left|\vec{c}_{12} \cdot \vec{e}\,\right| 
\phi\left(c_1\right) \phi\left(c_2\right) {\rm Expr}\left(\dots\right)\, ,
\end{equation} 
where $\text{Expr} \left(\dots\right)$  specifies a mathematical expression in terms of the pre- and aftercollision velocities, the Sonine polynomials and their coefficients and other variables (see below). The evaluation of such integrals by means of basic integrals $J_{k,l,m,n,p,\alpha}$ is straightforward, although for practical applications even simple looking expressions may result in sums of thousands of terms. Since the algebra follows an algorithm, we can use a formula manipulation program, such as Maple to evaluate kinetic integrals.

Before giving a systematic description we wish to describe the example from section \ref{sec:example} in detail, where we computed the second moment of the collision integral $\mu_2$ in linear approximation with respect to $a_2$. 

The kernel of the kinetic integral Eq. \eqref{mupa2x} reads 
\begin{equation}
  \label{eq:kernel}
\text{Expr}=-\frac12 \left\{1+a_2\left[S_2\left(c_1^2\right)+S_2\left(c_2^2\right) \right] \right\}\Delta \left(c_1^2+c_2^2\right)  
\end{equation}

The computation in the listing in section \ref{sec:example} is started with \verb|restart| to clean the memory of the computer from earlier results, then with \verb|libname...| we extend the library path of Maple and load the library (\verb|with...|). The next line is an output produced by maple, indicating that the commands \verb|DefDimension|, \verb|DefJ|, \verb|DefS|, \verb|KIinit|, \verb|getJexpr|, and \verb|unDefJ| can be used now. Here starts the evaluation of our integral: First we initialize the computation by \verb|KIinit()| which defines some internal variables. Then the dimension of the problem is specified. Here we wish to evaluate the Integral in three dimensions. With the line \verb|DefJ()| we require that the Basic Integrals will be evaluated, i.e., in the final result we generate real numbers according to \eqref{eq:Jn0eval4}-\eqref{eq:Jn2eval2}, e.g., $J_{0,0,2,0,2,0}=6\sqrt{2\pi}$. The line \verb|Expr:=...| defines the kernel of the integral due to Eq. \eqref{eq:App:intMaple}. Finally \verb|mu2:=getJexpr(0,Expr,0)| starts the computation. The arguments of \verb|getJexpr()| as well as the syntax of \verb|Expr| are described in detail below. 

We explain the Maple package \verb|KineticIntegral.m| in detail. To this end we comment the program line by line, bold numbers refer to line numbers in the listing. The package as well as the examples can be obtained at \cite{KI}. The program uses internal variables whose names follow a regular scheme: \verb|c1p|, \verb|c2p|, \verb|c1a|, and \verb|c2a| stand for the pre-collision and after-collision velocities of the particles. The unit vector \verb|e|$\equiv E\equiv\vec{r}_{12}/r_{12}$ and the dimensionality \verb|DIM| are further variables. Scalar products are named by \verb|xDOTy|, where \verb|x| and \verb|y| are vectors. An example is \verb|c1pDOTc2p|$\equiv  \vec{c}_1\cdot\vec{c}_2$. A full list of the internal variables is given in the table below. All these symbols can be used to specify the kernel \verb|Expr| of the kinetic integral.

\noindent \label{VarTable}\begin{tabular}{lll}
\verb|c1p| & $\vec{c}_1$:~~ pre-collision velocity of particle 1 \\
\verb|c2p| & $\vec{c}_2$:~~ pre-collision velocity of particle 2 \\
\verb|c1a| & $\vec{c}^{\,\prime}_1$:~~ after-collision velocity of particle 1 \\
\verb|c2a| & $\vec{c}^{\,\prime}_2$:~~ after-collision velocity of particle 2 \\
\verb|c12| & $\vec{c}_{12}$:~~ relative velocity\\
\verb|C| & $\vec{C}$:~~ centre of mass velocity\\
\verb|c12DOTe|=\verb|eDOTc12|& $\vec{c}_{12}\cdot \vec{e}$\\
\verb|CDOTc12|=\verb|c12DOTC|& $\vec{C}\cdot \vec{c}_{12}$\\ 
\verb|c1pDOTc1p| & $\left(\vec{c}_1\right)^2$ $ =\vec{C}^2+\vec{c}_{12}^{\,\,2}/4 + \vec{C}\cdot \vec{c}_{12}$ \\
\verb|c2pDOTc2p| & $\left(\vec{c}_2\right)^2$ $ =\vec{C}^2+\vec{c}_{12}^{\,\,2}/4 - \vec{C}\cdot \vec{c}_{12}$ \\
\verb|c1pDOTc2p| =\verb|c2pDOTc1p| &  $\vec{c}_1\cdot\vec{c}_2$ $ =\vec{C}^2-\vec{c}_{12}^{\,\,2}/4$\\
\verb|c1aDOTc1a| & $\left(\vec{c}^{\,\prime}_1\right)^2$ $=\vec{C}^2+\vec{c}_{12}^{\,\,2}/4-\left(1-\varepsilon^2\right)/4 \left(\vec{c}_{12}\cdot \vec{e}\,\right)^2 +$\\
                 & $~~~+ \vec{C}\cdot\vec{c}_{12}-(1+\varepsilon)\left(\vec{C}\cdot\vec{e}\,\right) \left(\vec{c}_{12}\cdot\vec{e}\,\right)$\\
\verb|c2aDOTc2a| & $\left(\vec{c}^{\,\prime}_2\right)^2$ $=\vec{C}^2+\vec{c}_{12}^{\,\,2}/4-\left(1-\varepsilon^2\right)/4 \left(\vec{c}_{12}\cdot \vec{e}\,\right)^2 -$\\
                 & $~~~- \vec{C}\cdot\vec{c}_{12}+(1+\varepsilon)\left(\vec{C}\cdot\vec{e}\right) \left(\vec{c}_{12}\cdot\vec{e}\,\right)$\\
\verb|c1aDOTc2a|=\verb|c2aDOTc1a| & $\vec{c}^{\,\prime}_1\cdot\vec{c}^{\,\prime}_2$  $=\vec{C}^2-\vec{c}_{12}^{\,\,2}/4+\left(1-\varepsilon^2\right)/4\left(\vec{c}_{12}\cdot \vec{e}\,\right)^2$ \\
\verb|c1pDOTc1a|=\verb|c1aDOTc1p| & $\vec{c}_1\cdot\vec{c}^{\,\prime}_1$  $=\vec{C}^2+\vec{c}_{12}^{\,\,2}/4+\vec{C}\cdot\vec{c}_{12}-$\\
                                  & $-(1+\varepsilon)/2 \left(\vec{C}\cdot\vec{e}\right) \left(\vec{c}_{12}\cdot\vec{e}\,\right)-(1+\varepsilon)/4\left(\vec{c}_{12}\cdot\vec{e}\,\right)^2$ \\
\verb|c2pDOTc2a|=\verb|c2aDOTc2p| & $\vec{c}_2\cdot \vec{c}^{\,\prime}_2$  $=\vec{C}^2+\vec{c}_{12}^{\,\,2}/4-\vec{C}\cdot\vec{c}_{12}+$\\
                                  & $+(1+\varepsilon)/2 \left(\vec{C}\cdot\vec{e}\right) \left(\vec{c}_{12}\cdot\vec{e}\,\right)-(1+\varepsilon)/4\left(\vec{c}_{12}\cdot\vec{e}\,\right)^2$ \\
\verb|c1aDOTc2p|=\verb|c2pDOTc1a| & $\vec{c}^{\,\prime}_1\cdot \vec{c}_2$  $=\vec{C}^2-\vec{c}_{12}^{\,\,2}/4-$\\
                                  & $-(1+\varepsilon)/2 \left(\vec{C}\cdot\vec{e}\right) \left(\vec{c}_{12}\cdot\vec{e}\,\right)+(1+\varepsilon)/4 \left(\vec{c}_{12}\cdot\vec{e}\,\right)^2$\\
\verb|c1pDOTc2a|=\verb|c2aDOTc1p| & $\vec{c}_1\cdot \vec{c}^{\,\prime}_2$  $=\vec{C}^2-\vec{c}_{12}^{\,\,2}/4+$\\
                                  & $+(1+\varepsilon)/2 \left(\vec{C}\cdot\vec{e}\right) \left(\vec{c}_{12}\cdot\vec{e}\,\right)+(1+\varepsilon)/4 \left(\vec{c}_{12}\cdot\vec{e}\,\right)^2$\\
\verb|Delta2|, \verb|Delta4| & see Eqs. \eqref{Deltac1c2p2}, \eqref{Deltac1c24}
\\
\verb|DIM| & dimension \\
\verb|C1|, \verb|dprime| & describes the coefficient of restitution of\\
                         & viscoelastic particles, see \eqref{epsc12} !!!! \\
\verb|absc12DOTe5| & $\left|\vec{c}_{12}\cdot \vec{e}\,\right|^{1/5}$
\end{tabular}

\medskip

\noindent
The variables \verb|Delta2| and \verb|Delta4| are defined in the initialization \verb|KIinit()|:

\newprog
\proginput{KineticIntegralPackageDim.txt}{KIinit}

\noindent
The kernel \verb|Expr| may contain Sonine polynomials. They are defined by \verb|DefS()| according to Eq. \eqref{eq:Soninedef}. 

\proginput{KineticIntegralPackageDim.txt}{DefS}

\noindent
First at line {\bf 14} it is checked whether the dimensionality \verb|DIM| is specified, otherwise \verb|DefS| reports an error at line {\bf 38}. The syntax of the Sonine polynomials is \verb|S(a,expr)|, \verb|a| specifies the order of the polynomial and \verb|expr| is its algebraic argument which may contain the pre- and after-collision particle velocities \verb|c1p|, \verb|c1a|, \verb|c2p|, and \verb|c2a|. In lines {\bf 18-21} these velocities are translated into expressions of the center of mass velocity \verb|C|, the relative velocity \verb|c12|, the coefficient of restitution \verb|epsilon|, the unit vector \verb|e| and the scalar product \verb|c12DOTe|, due to Eq. \eqref{directcollviaCc}. There is an important difference in the notations of the Sonine coefficients due to Eq. \eqref{eq:Soninedef} and in the Maple program:
\begin{equation}
\verb|S(a,expr)| \equiv S_a \left({\rm expr}^2\right)\,.  
\end{equation}
This difference of the notation is taken into account by line {\bf 22} where the entire term \verb|Expr| with all the described substitutions is squared. In the squared expressions there appears the term $\vec{C}\cdot\vec{c}_{12}$ which is again substituted by \verb|CDOTc12|. The program evaluates Sonine polynomials of the order $a=\{0,1,2\}$ which is done in line {\bf 24-29}. Any incorrect input \verb|a| is acknowledged by an error message, line {\bf 31}. 

All of the defined internal variables, including the Sonine polynomials are accessible within the Maple program. In general, it is of not much value to list them since according to the substitution of variables even small terms become impressive expressions. For debugging purposes, however, this feature may be useful.

There are two different output modes: first we may be interested in the result of an integration in terms of basic integrals $J_{k,l,m,n,p,\alpha}$ and second (usually) we are interested in obtaining real numbers. The translation of the basic integrals into real numbers is given by Eqs. \eqref{eq:Jn0eval4}-\eqref{eq:Jn2eval2}. These equations are implemented in \verb|DefJ()|:

\proginput{KineticIntegralPackageDim.txt}{DefJ}

\noindent
 After calling \verb|defJ| all results are given as real numbers by evaluating the basic integrals, e.g.,
\medskip

\noindent\hspace*{0.cm}\fbox{\begin{minipage}{11.5cm}
{\footnotesize 
 
\DefineParaStyle{Maple Output}
\DefineCharStyle{2D Math}
\DefineCharStyle{2D Output}



%





\begin{maplegroup}
\begin{mapleinput}
\mapleinline{active}{1d}{DefJ();}{%
}
\end{mapleinput}

\mapleresult
\begin{maplettyout}
Basic Integrals will be evaluated
\end{maplettyout}

\end{maplegroup}
\begin{maplegroup}
\begin{mapleinput}
\mapleinline{active}{1d}{BIdemo:=J(0,0,1,1,1,1);}{%
}
\end{mapleinput}

\mapleresult
\begin{maplelatex}
\mapleinline{inert}{2d}{BIdemo := 11/100*2^(3/5)*Pi^(3/2)/sin(1/10*Pi)/GAMMA(9/10);}{%
\[
\mathit{BIdemo} := {\displaystyle \frac {11}{100}} \,
{\displaystyle \frac {2^{(3/5)}\,\pi ^{(3/2)}}{\mathrm{sin}(
{\displaystyle \frac {1}{10}} \,\pi )\,\Gamma ({\displaystyle 
\frac {9}{10}} )}} 
\]
}
\end{maplelatex}

\end{maplegroup}
\begin{maplegroup}
\begin{mapleinput}
\mapleinline{active}{1d}{evalf(BIdemo);}{%
}
\end{mapleinput}

\mapleresult
\begin{maplelatex}
\mapleinline{inert}{2d}{2.811423645;}{%
\[
2.811423645
\]
}
\end{maplelatex}

\end{maplegroup}

}
\end{minipage}}
\medskip

\noindent The counterpart of \verb|DefJ()| is \verb|unDefJ()|. After calling \verb|unDefJ| any result is given in terms of basic integrals.

\proginput{KineticIntegralPackageDim.txt}{unDefJ}
\noindent
In the described programs \verb|DefS| and \verb|DefJ| we have already used the variable \verb|DIM| which contains the dimensionality of the Basic Integral. This variable is set by \verb|DefDimension()|:

\proginput{KineticIntegralPackageDim.txt}{DefDimension}

\noindent
After changing the variable \verb|DIM| in line {\bf 78} the Sonine polynomials which are specific for any dimension have to be redefined (line {\bf 81}). For the case that \verb|DefJ()| has been called before, the basic integrals have to be redefined as well, line {\bf 83}.

The heart of the Maple package is the program \verb|getJexpr()|. This program has three arguments: \verb|epstype| specifies the type of the particle interaction. \verb|epstype|=0 stands for constant coefficient of restitution and \verb|epstype|=1 stands for viscoelastic particles. The second argument is the kernel of the integral \verb|expr| which is a function of the particle velocities before and after the collision, the dimension, the Sonine polynomials and their arguments, the Sonine coefficients, and scalar products of the vectorial values. In fact \verb|expr| may contain all variables which are listed in the table on page \pageref{VarTable}. The third argument describes the output mode of the program. It will be described later.
The text of the program \verb|getJexpr()| is listed below:

\proginput{KineticIntegralPackageDim.txt}{getJexpr}

\noindent
First, in lines {\bf 91-112} the kernel of the integral \verb|expr| is processed. All variables are expressed in terms of \verb|epsilon|, \verb|C|, \verb|c12|, \verb|CDOTc12|, \verb|CDOTe|, \verb|c12DOTe|, \verb|a2|. If we deal with viscoelastic particles (\verb|epstype|=1), the coefficient of restitution is a function of the relative velocity. It is expressed in our variables by Eq. \eqref{epsc12}, where the coefficients $\delta^\prime$ describes the material properties of the viscoelastic particles and $C_1$ is a numerical constant, see Eq. \eqref{epsinv}. Therefore, for this case $\varepsilon$ has to be replaced by an expression in our variables which is done in lines {\bf 116-117}. We introduce a new internal variable \verb|absc12DOTe5|$\equiv \left|\vec{c}_{12} \cdot \vec{e}\, \right|^{1/5}$.

The entire kernel of the integral is then in line {\bf 121} represented as a sum of polynomials in the variables  \verb|C|, \verb|c12|, \verb|CDOTc12|, \verb|CDOTe|, \verb|c12DOTe|, \verb|a2|, \verb|epsilon| (for $\varepsilon=\mbox{const.}$), and \verb|absc12DOTe5| (for viscoelastic particles). These are precisely the variables which define our basic integrals for which we know the solution (see Eqs. \eqref{eq:Jn0eval4}-\eqref{eq:Jn2eval2}). For any of these terms we determine in the loop (lines {\bf 125-142}) the indices of the according basic integral. The terms are then successively picked out of the sum by line {\bf 126} and stored in \verb|term|. The next line determines the coefficients of the variables in this term, they are stored temporarily in the list \verb|coeffs|. In lines {\bf 128-134} the powers of the variables in the present term are determined, i.e., \verb|epow| is the power of \verb|epsilon| in \verb |term|, \verb|aa| is the power of \verb|absc12DOTe5|, etc. These powers determine the indices of the basic integral. 
In this way all terms of the sum are transformed in line {\bf 136} into expressions of the basic integrals and the according prefactors. By processing the terms in the loop the variable \verb|termJ| grows and finally it contains the full sum in terms of basic integrals. 

The vector \verb|di[]| in line {\bf 137} is for testing purposes. It reconstructs the present term by using the powers of the variables as computed above and its prefactors. If the reconstructed term differs from the original term (line {\bf 139}) an error has occurred. This may happen by an typing error for the kernel of the integral. The program then reports the error and the number of the concerned term.

In line {\bf 143} the integral is completely expressed in terms of basic integrals. There are different output modes, specified by the third argument \verb|ou| of the program \verb|getJexpr()|. If \verb|ou=0| the output is given in symbolic form, using the Gamma-function and its properties. The result is analytically exact, however, the output may become very large. For \verb|ou=1| the result is given by a floating point number. Before, the numerical constant $C_1$ is replaced in line {\bf 146} by its analytical expression Eq. \eqref{eq:defC1}.
For the choice \verb|ou=2| the result is given as a Taylor expansion using the numerical constants $\omega_{1/2}$, which  make the expression more compact. These constants read:
\begin{equation}
  \label{eq:omegadef}
\omega_0 \equiv  2 \sqrt{2 \pi} 2^{1/10} \Gamma \left (\frac{21}{10} \right)C_1\approx 6.485\,,\qquad
\omega_1 \equiv  \sqrt{2 \pi} 2^{1/5} \Gamma \left (\frac{16}{5} \right)C_1^2\approx 9.285\,.
\end{equation}
 Finally the defined commands have to be bundled to a package and stored with the name \verb|KineticIntegral.m|. This is done by the last lines of the program:
\proginput{KineticIntegralPackageDim.txt}{makePackage}
The presented program generates the maple library \verb|KineticIntegral.m|. To evaluate integrals of the type Eq. \eqref{eq:App:intMaple} the library can be invoked and used as demonstrated in the example in section \ref{sec:example}.

\section{Kinetic integrals in the kinetic theory of Granular Gases}
\subsection{Homogeneous cooling state}
\subsubsection{Temperature decay and velocity distribution function for the simplified collision model ($\varepsilon=\text{const.}$)}

In the first stage of the evolution of an initially homogeneous gas of dissipative particles the gas preserves its homogeneity, however, its temperature decays persistently. This state of the evolution is called the homogeneous cooling state. In this state the velocity distribution function does not depend on the space coordinate $\vec{r}$. Hence, the homogeneous cooling state is characterized by the gas temperature $T$ as defined by Eq. \eqref{deftemp1}.

In our introducing example in section \ref{sec:example} we have derived the evolution of temperature of a Granular Gas. The deviation of the velocity distribution function from the Maxwell distribution is characterized by the Sonine coefficients $a_i$ (see Eq. \eqref{Soninexp}). If the deviations from the Maxwell distributions are small only the first nonvanishing term in the expansion Eq. \eqref{Soninexp} contributes considerably to the distribution function, i.e. we use approximation (\ref{fa2x}). For the simplified collision model ($\varepsilon =\mbox{const.}$) the second Sonine coefficient $a_2$ is related with the moments of the collision integral by \cite{NoijeErnst:1998,BrilliantovPoeschelStability:2000}
\begin{equation}
\label{eqa2}
5\mu_2 \left(1+a_2 \right)-\mu_4 =0\, .
\end{equation} 
This equation is solved by the program below:

\newprog
{\scriptsize 
\begin{Verbatim}
DefDimension(3);
DefJ();
Expr:=Delta2*(-1/2*(1+a2*(S(2,c1p)+S(2,c2p)) + a2^2*S(2,c1p)*S(2,c2p))):
mu2:=getJexpr(0,Expr,0);
Expr:=Delta4*(-1/2*(1+a2*(S(2,c1p)+S(2,c2p)) + a2^2*S(2,c1p)*S(2,c2p))):
mu4:=getJexpr(0,Expr,0);
A2:=solve(5*mu2*(1+a2)-mu4=0,a2):
AA2:=unapply(Re(A2[2]),epsilon):
A2NE:=epsilon->(16*(1-epsilon)*(1-2*epsilon^2))
  /(81-17*epsilon+30*epsilon^2*(1-epsilon)):
plot([AA2,A2NE],0..1,linestyle=[1,4],
 legend=["nonlin.theory","lin.theory (v. Noije, Ernst)"],
 font=[COURIER,20],thickness=2,labels=[e,a2],color=black);
\end{Verbatim}
}

\noindent
As a result we obtain the second Sonine coefficient as a function of the coefficient of restitution as drawn in Fig. \ref{fig:a2eps}.
\begin{figure}[htbp]
  \centerline{\includegraphics[width=6cm,height=5cm]{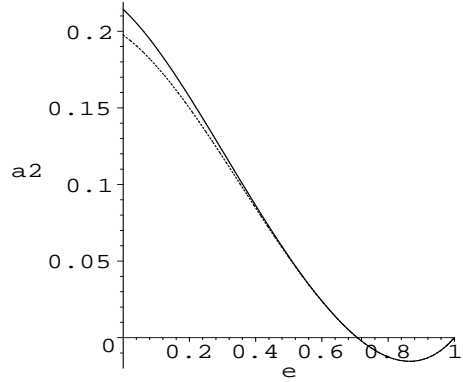}}
  \caption{Second Sonine coefficient $a_2$ over the coefficien of restitution $\varepsilon$. Full line: solution of Eq. \eqref{eqa2} (nonlinear theory \cite{BrilliantovPoeschelStability:2000}), dashed line: linear solution  of Eq. \eqref{eqa2}, i.e. only linear terms with respect to $a_2$ are kept in this equation \cite{NoijeErnst:1998}.}
  \label{fig:a2eps}
\end{figure}
The full analytic expression for $a_2(\varepsilon)$ can be found in \cite{BrilliantovPoeschelStability:2000}.

\subsubsection{Temperature decay and velocity distribution function for gases of viscoelastic particles}

For Granular Gases of viscoelastic particles under the adiabatic approximation \cite{BrilliantovPoeschel:2003} the cooling rate and the second Sonine coefficient are computed in the same way as for $\varepsilon=\text{const.}$ The according program reads
\label{page:mu2mu4visco}
\newprog
{\scriptsize 
\begin{Verbatim}
DefDimension(3):
DefJ():
Expr:=Delta2*(-1/2*(1+a2*(S(2,c1p)+S(2,c2p)) + a2^2*S(2,c1p)*S(2,c2p))):
mu2:=getJexpr(1,Expr,2);
Expr:=Delta4*(-1/2*(1+a2*(S(2,c1p)+S(2,c2p)) + a2^2*S(2,c1p)*S(2,c2p))):
mu4:=getJexpr(1,Expr,2):
A2:=solve(5*mu2*(1+a2)-mu4=0,a2);
\end{Verbatim}
}

\noindent
which yields $\mu_2$ on line {\bf 5} in terms of $a_2$ and then $a_2$ on line {\bf 9}. Note that on line {\bf 9} we need to take a physical root $a_2 \propto \delta^{\prime}$.  The result for $a_2$ reads
\begin{equation}
\label{a2adiab}
a_2    = a_{21} \delta^{\, \prime}  + a_{22} \delta^{\, \prime \, 2}  + \dots 
\end{equation}
\begin{equation}
a_{21} = -\frac{3  \omega_0}{20 \sqrt{2 \pi}} \approx  -0.388\,,\qquad
a_{22} = \frac{12063}{640000} \frac{\omega_0^2}{\pi} + \frac{27}{40} \frac{\omega_1}{\sqrt{2 \pi}} \approx 2.752 \, .
\end{equation}
Correspondingly the cooling coefficient, expressed via $\mu_2$ (see Eq. \eqref{eq:zeta_def_hom1}) reads
\begin{equation}
\label{zetavisco}
\zeta = \frac23 \sqrt{\frac{2T}{m}} n \sigma^2   \left( \omega_0 \delta^{\, \prime} - \omega_2 \delta^{\, \prime \, 2}+\dots \right)\, ,
\end{equation}
where 
\begin{equation}
\omega_2 \equiv \omega_1+ \frac{9}{500}\omega_0^2 \sqrt{\frac{2}{\pi}}\, ,   
\end{equation}
which yields for the temperature evolution in linear approximation with respect to $\delta$ 
\begin{equation}
\label{T(t)del1}
\frac{T(t)}{T_0}= \left(1 + \frac{t}{\tau_0} \right)^{-5/3}
\end{equation}
where we introduce the characteristic time
\begin{equation}
\label{tau0}
\tau_0^{-1}=\frac{16}5 q_0 \delta \tau_c(0)^{-1}= \frac{48}{5}q_0 \delta  \,4 \sigma^2 n\sqrt{\frac{\pi T_0}{m}} \, ,
\end{equation}
with the initial mean collision time $\tau_c(0)$ and the constant 
\begin{equation}
q_0=2^{1/5}\Gamma\left(\frac{21}{10}\right) \frac{C_1}{8} \approx 0.173\,.   
\end{equation}
The evaluation of $a_2$ requires already the solution of 786 basic integrals.
The above results for the temperature evolution and second Sonine coefficient have been reported in \cite{BrilliantovPoeschelStability:2000,BrilliantovPoeschel:2000visc}.

\subsubsection{Coefficient of self-diffusion for the simplified collision model ($\varepsilon=\text{const.}$)}

If we assume Molecular Chaos in the homogeneous cooling state the velocity correlation function  
\begin{equation}
\left< \vec{v}(t) \vec{v}(t^{\prime}) \right> = \left< \vec{v}(t^{\prime})^2 \right> \exp\left(-\frac{\left|t-t^{\prime}\right|}{\tau_v}\right)  
\end{equation}
is related with the coefficient of self-diffusion 
\begin{equation}
D(t)  = \frac{T}{m} \tau_{v}(t)
\label{eq:D_byond_ad}
\end{equation} 
via the velocity relaxation time $\tau_v$. This quantity reads \cite{BrilliantovPoeschel:1998d}
\begin{multline}
  \label{eq:tauvgen2}
  \tau_v^{-1}(t) 
  =   \frac{1}{12 \sqrt{2 \pi}} \tau_c^{-1}(t) \\
  \times \int d\,\vec{c}_1 d\,\vec{c}_2 \int d\vec{e}\, \Theta\left(-\vec{c}_{12} \cdot \vec{e}\,\right) \left| \vec{c}_{12} \cdot \vec{e}\, \right| 
  \tilde{f}\left(\vec{c}_1, t\right)\tilde{f}\left(\vec{c}_2, t\right) \left( 1 + \varepsilon \right) \left( \vec{c}_{12} \cdot \vec{e}\, \right)^2 \,. 
\end{multline}
Again we recognize the kinetic integral in Eq. \eqref{eq:tauvgen2} which allows for the application of the Maple package. Using the program
\newprog
{\scriptsize 
\begin{Verbatim}
DefDimension(3);
unDefJ();
Expr:=tau_c_inv/(12*sqrt(2*Pi))*(1+epsilon)*c12DOTe*c12DOTe*(1+a2*S(2,c1p))
      *(1+a2*+S(2,c2p))):
DefJ();
tau_v_inv:=getJexpr(0,Expr,0);
\end{Verbatim}
}
\noindent
we obtain the velocity relaxation time
\begin{equation}
\tau_v^{-1}(t) = \frac{\left(1 +\varepsilon \right) }{3} \left( 1+ \frac{3a_2}{32} \right)^2 \tau_c^{-1}(t)
\end{equation} 
which yields with Eqs. (\ref{eq:D_byond_ad}) the self-diffusion coefficient 
\begin{equation}
D(t)= \frac{2D_0(t)}{ \left( 1+ \varepsilon  \right) \, \left( 1 + \frac{3}{32}a_2 \right)^2}  \, , 
\label{a2Difviatau_sol}
\end{equation}
where 
\begin{equation}
\label{DEnskog}
D_0^{-1}(t)= \frac83 \sigma^2  n  \sqrt{ \frac{\pi m}{T(t) }} \, .
\end{equation}
is the Enskog self-diffusion coefficient of the elastic gas. 

The self-diffusion coefficient in the linear approximation with respect to $a_2$ has been computed for the simplified collision model in \cite{BreyRuizMonteroCuberoGarcia:2000}.

\subsubsection{Coefficient of self-diffusion for gases of viscoelastic particles}

A similar analysis can be performed for gases of viscoelastic particles:

\newprog
{\scriptsize 
\begin{Verbatim}
DefDimension(3):
DefJ():
Expr:=(1+epsilon)*c12DOTe*c12DOTe*(1+a2*(S(2,c1p)+S(2,c2p))+a2^2*S(2,c1p)*S(2,c2p)):
hattauV:= getJexpr(1,Expr,2):
tauVinv:=1/(2*DIM)*sqrt(2*T/m)*n*sigma^(DIM-1)*hattauV:
Expr:=Delta2*(-1/2*(1+a2*(S(2,c1p)+S(2,c2p))+a2^2*S(2,c1p)*S(2,c2p))):
mu2:=getJexpr(1,Expr,2):
Expr:=Delta4*(-1/2*(1+a2*(S(2,c1p)+S(2,c2p))+a2^2*S(2,c1p)*S(2,c2p))):
mu4:=getJexpr(1,Expr,2):
tmpA2:={solve(mu4=mu2/DIM*4*DIM/4*(DIM+2)*(1+a2),a2)}:
a2:=tmpA2[1]:
a2:=convert(a2,polynom):
tauVinv:=convert(taylor(tauVinv,dprime,3),polynom):
Dvis:=T/m/tauVinv:
DEnsk:=DIM*GAMMA(DIM/2)*sqrt(T/m) / (4*Pi^((DIM-1)/2)*n*sigma^(DIM-1)):
simplify(convert(taylor(Dvis/DEnsk,dprime,3),polynom));
\end{Verbatim}
}

\noindent
At line {\bf 14} we need to take a root $a_2 \propto \delta^{\prime}$. 
The coefficient of self-diffusion for gases of viscoelastic particles in linear approximation with respect to $a_2$ has been reportet in \cite{BrilliantovPoeschel:1998d}. Here we give the complete solution with respect to $a_2$ which becomes feasible due to symbolic algebra:
\begin{equation}
D(t)  = D_0(t)\left[ 1 + \frac{89}{640} \sqrt{\frac{2}{\pi}} \omega_0 \, 
\delta^{\prime} + \left( \frac{625537}{20480000} \frac{\omega_0^2}{\pi} 
 - \frac{1851}{14080} \sqrt{\frac{2}{\pi}} \omega_1 \right) \delta^{\prime \, 2}  + \cdots \right]  
\label{eq:D_expand_final}
\end{equation} 

\subsection{Inhomogeneous Granular Gases}
\subsubsection{Transport coefficients}

So far we have considered Granular Gases in the homogeneous cooling state. To describe the non-homogeneous gas a number of perturbative approaches has been developed. The most successive are the Chapman-Enskog approach \cite{ChapmanCowling:1970,FerzigerKaper:1972} and the Grad method \cite{Grad:1960}; here we consider the former one. 

In the Chapman-Enskog approach it is assumed that (after a short relaxation time) the time and space dependence of the velocity distribution function $f(\vec{r},\vec{v},t)$ occurs only via  density $n(\vec{r},t)$, flow velocity $\vec{u}( \vec{r},t )$ and temperature $T(\vec{r},t)$. These quantities are, respectively, the zeroth, first and second order moments of the velocity distribution function; $n(\vec{r},t)$  and $\vec{u}( \vec{r},t )$ are expressed in terms of  $f(\vec{r},\vec{v},t)$ in the same way is temperature in \eqref{deftemp1} \cite{ChapmanCowling:1970,FerzigerKaper:1972}. These three fields evolve according to the hydrodynamic equations (e.g. \cite{BrilliantovPoeschel:2001roy,BrilliantovPoeschel:2003}), 
\begin{equation}
\label{eq:hydroeq_sum}
\begin{split}
    \frac{\partial n}{\partial t} + \vec{\nabla} \cdot \left( n \vec{u} \right) & =0 \\
    \frac{\partial \vec{u}}{\partial t} + \vec{u} \cdot \vec{\nabla} \vec{u} + \left(nm \right)^{-1} \vec{\nabla} \cdot \hat {P} &=0 \\
    \frac{\partial T}{\partial t} + \vec{u} \cdot \vec{\nabla} T + \frac{2}{3n} \left( P_{ij} \nabla_j u_i + \vec{\nabla} \cdot \vec{q} \right) + \zeta \, T &=0  \, ,  
\end{split}
\end{equation}
where $\hat {P}$ denotes the pressure tensor and $\vec{q}$ is the vector of the heat flux. In the Navier-Stokes hydrodynamics the pressure tensor and the heat flux depend linearly on the field gradients: 
\begin{equation}
\label{eq:Pres_q_sum}
\begin{split}
    P_{ij}  & = p\delta_{ij} - \eta \left( \nabla_i u_j +\nabla_j u_i -\frac{2}{3} \delta_{ij} \vec{\nabla} \cdot \vec{u} \right)  \\
    \vec{q} & = -\kappa \vec{\nabla} T - \mu \vec{\nabla} n   \, ,
\end{split}
\end{equation}
with the hydrostatic pressure $p=nT$ in the dilute limit. Here $\eta$ is the viscosity coefficient, $\kappa$ is the coefficient of thermal conductivity and coefficient $\mu$ characterises the heat flux due to the density gradients and does have an analogue for elastic gases. 


The transport coefficients $\eta$, $\kappa$, $\mu$ may be calculated using the Chapman-Enskog approach adopted for the dissipative gases. The most complete analysis of this problem for the gas with a simplified collisional model with a constant $\varepsilon$ have  been performed by Brey and coauthors \cite{BreyDuftyKimSantos:1998,BreyCubero:2000}.  They have expressed transport coefficients for the dissipative gases in terms of functions of the collision integral. The generalization of this approach for the dissipative gases with the velocity-dependent restitution coefficient has been done in \cite{BrilliantovPoeschel:2003}, where it has been shown that  the transport coefficient may be expressed in terms of the same functions of the collision integral as for the case of $\varepsilon =\mbox{const.}$

\subsubsection{Transport coefficients for the simplified collision model}
Using the notations of \cite{BrilliantovPoeschel:2003} the viscosity coefficient reads for the case of $\varepsilon=\mbox{const.}$ (see also \cite{BreyDuftyKimSantos:1998}),  
\begin{equation}
\label{eq:eta_via_mu2Omega}
\eta=-\frac{1}{\sigma^2 } \sqrt{\frac{mT}{2}} \frac{1}{ \frac13 \mu_2 + \frac25 \Omega_{\eta}} \, .
\end{equation} 
where $\mu_2$ has been found above (see Eq. \eqref{mu2A}) and with the function of the collision integral $\Omega_{\eta}$  
\begin{multline}
\label{eq:def_Omegaeta_gen}
\Omega_{\eta} \equiv \int d \vec{c}_1 \int d \vec{c}_2 \int d\vec{e}\, 
        \Theta\left(-\vec{c}_{12} \cdot \vec{e}\,\right) \left|\vec{c}_{12} \cdot \vec{e}\,\right|  
        \tilde{f}^{(0)}(c_1)  \times \\
\times \phi(c_2) D_{ij}(\vec{c}_2)\Delta \left[D_{ij}(\vec{c}_1) + D_{ij}(\vec{c}_2) \right] \, ,  
\end{multline}
where $D_{ij}\left(\vec{c}\,\right)$ is the traceless tensor, 
\begin{equation}
\label{eq:defDcij}
D_{ij} \left(\vec{c}\,\right) \equiv  c_ic_j -\frac13 \delta_{ij}c^2  \, . 
\end{equation}

In Eq. \eqref{eq:def_Omegaeta_gen}  we recognize the kinetic integral. For the simplified collision model ($\varepsilon=\text{const.}$) it is solved by the program

\newprog
{\scriptsize 
\begin{Verbatim}
DefDimension(3);
DefJ();
Expr :=(-1/DIM*c2pDOTc2p*Delta2+c1aDOTc2p*c1aDOTc2p+c2aDOTc2p*c2aDOTc2p
       -c2pDOTc2p*c2pDOTc2p-c1pDOTc2p*c1pDOTc2p)*(1+a2*S(2,c1p)): 
OmegaETA:=getJexpr(0,Expr,0);
\end{Verbatim}
}

\noindent
yielding
\begin{equation}
\label{eq:Omegaeta_eps_Maple}
\Omega_{\eta} = -\sqrt{2 \pi} \left( 1 + \varepsilon \right) \left( 3 - \varepsilon \right) \left( 1-\frac{a_2}{32} \right) \, ,  
\end{equation} 

Similarly the coefficient of thermal conductivity may be expressed as 
\begin{equation}
\label{eq:kapOmmu2_eps}
\kappa=-\frac{15}{8 \sigma^2 g_2(\sigma) } \sqrt{\frac{T}{2 m}} \frac{ 1 + 2 a_2 }{ \mu_2 +\frac15 \Omega_{\kappa}}\, ,  
\end{equation}
with the function $ \Omega_{\kappa}$ defined as 
\begin{multline}
\label{eq:def_Omegakap_gen}
 \Omega_{\kappa} \equiv 
    \int d \vec{c}_1 \int d \vec{c}_2 \int d\vec{e}\, \Theta\left(-\vec{c}_{12} \cdot \vec{e}\,\right)\left|\vec{c}_{12} \cdot \vec{e}\,\right| \tilde{f}^{(0)}(c_1)   \times  \\
\times \phi(c_2) \vec{S} \left( \vec{c}_2  \right)  \cdot  \Delta \left[ \vec{S} \left( \vec{c}_1 \right) + \vec{S} \left( \vec{c}_2 \right) \right] \, ,  
\end{multline}
and with the vectorial quantity, 
\begin{equation}
\label{eq:defS_c}
\vec{S}(\vec{c}) \equiv  \left( c^2 - \frac52 \right) \vec{c}  \, . 
\end{equation}
The coefficient $\mu$ may be also written in terms of $ \Omega_{\kappa}$:
\begin{equation}
\label{eq:mu_viaOmega}
\mu=\frac{5}{2 n \sigma^2} \sqrt{\frac{T}{2 m} } 
\frac{a_2}{\mu_2+\frac{4}{15}\Omega_{\kappa}}\, .  
\end{equation}

The function $\Omega_{\kappa}$ is computed by the program

\newprog
{\scriptsize 
\begin{Verbatim}
DefDimension(3);
DefJ();
Expr:=(c2pDOTc2p-(DIM+2)/2)*(c1aDOTc2p*c1aDOTc1a+c2aDOTc2p*c2aDOTc2a
      -c2pDOTc2p*c2pDOTc2p-c1pDOTc2p*c1pDOTc1p)*(1+a2*S(2,c1p)):
OmegaKAPPA:=getJexpr(0,Expr,0);
\end{Verbatim}
}

\noindent
We obtain
\begin{equation}
\label{eq:Omega_kap_eps}
\Omega_{\kappa}  
= -\sqrt{2 \pi} (1+\varepsilon) \left(\frac{49-33 \varepsilon}{8}+\frac{19- 3 \varepsilon}{256} a_2 \right) \,.  
\end{equation}
In linear approximation with respect to $a_2$ the transport coefficients have been reported in \cite{BreyDuftyKimSantos:1998}. Substituting Eqs. (\ref{eq:Omegaeta_eps_Maple},\ref{eq:Omega_kap_eps},\ref{mu2A}) into Eqs. (\ref{eq:eta_via_mu2Omega},\ref{eq:kapOmmu2_eps},\ref{eq:mu_viaOmega}) we obtain the generalized result for the coefficients $\eta$, $\kappa$ and $\mu$, which accounts for the terms of the next order, ${\mathcal O} \left( a_2^2 \right)$. 

\subsubsection{Transport coefficients for Granular Gases of viscoelastic particles}

The same function $\Omega_{\eta}$  appears in the theory of Granular Gases of viscoelastic particles \cite{BrilliantovPoeschel:2003}. We evaluate this function by

\newprog
{\scriptsize 
\begin{Verbatim}
DefDimension(3);
DefJ();
Expr:=(-1/DIM*c2pDOTc2p*Delta2+c1aDOTc2p*c1aDOTc2p+c2aDOTc2p*c2aDOTc2p
      -c2pDOTc2p*c2pDOTc2p-c1pDOTc2p*c1pDOTc2p)*(1+a2*S(2,c1p)):
OmegaETA:=getJexpr(1,Expr,2);
\end{Verbatim}
}

\noindent
and obtain the corresponding coefficient
\begin{equation}
\label{eq:Omegaeta_vis}
\Omega_{\eta}^{\rm vis} =  
-\left( w_0 + \delta^{\prime} w_1 - \delta^{\prime \, 2} w_2 \right)
\end{equation}
with 
\begin{equation}
  \begin{split}
    w_0&= 4  \sqrt{2 \pi} \left(1-\frac{1}{32}a_2 \right) \\
    w_1&= \omega_0 \left(\frac{1}{15}-\frac{1}{500}a_2 \right) \\
    w_2&= \omega_1 \left(\frac{97}{165}-\frac{679}{44000}a_2 \right) \, . 
  \end{split}
\end{equation}
Similarly, by

\newprog
{\scriptsize 
\begin{Verbatim}
DefDimension(3);
DefJ();
Expr:=(c2pDOTc2p-(DIM+2)/2)*(c1aDOTc2p*c1aDOTc1a+c2aDOTc2p*c2aDOTc2a
      -c2pDOTc2p*c2pDOTc2p-c1pDOTc2p*c1pDOTc1p)*(1+a2*S(2,c1p)):
OmegaKAPPA:=getJexpr(1,Expr,2);
\end{Verbatim}
}

\noindent
we obtain the other function
\begin{equation}
\label{eq:Omegakappa_vis}
\Omega_{\kappa}^{\rm vis} =  
-\left( u_0 + \delta^{\prime} u_1 - \delta^{\prime \, 2} u_2 \right)
\end{equation}
with 
\begin{equation}
\label{eq:u0u1u2_kappa_vis}
\begin{split}
u_0&=4\sqrt{2 \pi} \left(1+\frac{1}{32}a_2 \right) \\
u_1&=\omega_0 \left( \frac{17}{5}- \frac{9}{500}a_2 \right) \\
u_2&=\omega_1 \left( \frac{1817}{440}- \frac{1113}{352000} a_2 \right) \, .  
\end{split}
\end{equation}  

For the gas of viscoelastic particles the viscosity coefficient $\eta$ is found as the solution of the equation \cite{BrilliantovPoeschel:2003}  
\begin{equation}
\label{eq:for_etaviaOmega}
-\zeta T \frac{\partial \eta}{\partial T}=\frac25 \eta n \sigma^2 \sqrt{\frac{2T}{m}} \Omega_{\eta}^{\rm vis} +nT \, ,  
\end{equation}
where $\zeta$ is given by \eqref{zetavisco}. We solve this equation as a perturbative expansion with respect to the dissipative parameter $\delta^{\prime}$,
\begin{equation}
\label{etexpvisco}
\eta = \eta_0\left(1 + \delta^{\prime} \tilde{\eta}_1 + \delta^{\prime \, 2} \tilde{\eta}_2 + \cdots \right)\,.
\end{equation}
using the program

\newprog
{\scriptsize 
\begin{Verbatim}
DefDimension(3):
DefJ():
Expr:=Delta2*(-1/2*(1+a2*(S(2,c1p)+S(2,c2p))+a2^2*S(2,c1p)*S(2,c2p))):
mu2:=getJexpr(1,Expr,2):
mu2:=convert(mu2,polynom):
Expr:=Delta4*(-1/2*(1+a2*(S(2,c1p)+S(2,c2p))+a2^2*S(2,c1p)*S(2,c2p))):
mu4:=getJexpr(1,Expr,2):
tmpA2:={solve(mu4=mu2/DIM*4*DIM/4*(DIM+2)*(1+a2),a2)}:
a2:=tmpA2[1]:
a2:=convert(a2,polynom):
xi0:=2*sigma^(DIM-1)/DIM*n*mu2*sqrt(2*T/m):
ExprETA:=(-1/DIM*c2pDOTc2p*Delta2+c1aDOTc2p*c1aDOTc2p+c2aDOTc2p*c2aDOTc2p
 -c2pDOTc2p*c2pDOTc2p-c1pDOTc2p*c1pDOTc2p)*(1+a2*S(2,c1p)):
OmegaETA:=getJexpr(1,ExprETA,2):
OmegaETA:=convert(OmegaETA,polynom):
EqETA:=xi0*eta0*(1/2+3/5*dprime*eta1+7/10*dprime^2*eta2)
 +4/(DIM-1)/(DIM+2)*eta0*(1+dprime*eta1
 +dprime^2*eta2)*n*sigma^(DIM-1)*sqrt(2*T/m)*OmegaETA+n*T:
EqETA:=taylor(EqETA,dprime,3):
eta0:=solve(coeff(EqETA,dprime,0),eta0);
eta1:=solve(coeff(EqETA,dprime,1),eta1);
eta2:=solve(coeff(EqETA,dprime,2),eta2);
omega1:=9.28569:
omega0:=6.48562:
evalf(eta1);
evalf(eta2);
etaVIS:=eta0*(1+dprime*eta1+dprime^2*eta2);
plot(etaVIS/eta0,dprime=0..0.5,color=black,linestyle=[1]);
\end{Verbatim}
}

\noindent
The result reads
\begin{equation}
\label{eq:eta_fin}
\eta = \frac{5}{16 \sigma^2} \sqrt{\frac{mT}{\pi}} \left(1 + 0.483\, \delta^{\prime} + 0.094\, \delta^{\prime \, 2} + \dots \right)\,.
\end{equation}

Similarly, the coefficient of thermal conductivity may be obtained from the equation \cite{BrilliantovPoeschel:2003}, 
\begin{equation}
\label{eq:kappaviaOm}
-3 \frac{\partial }{\partial T} \zeta \kappa T = \frac45 \kappa n \sigma^2 \sqrt{\frac{2T}{m}} \Omega_{\kappa}^{\rm vis} +\frac{15}{4} \frac{nT}{m} \left( 1 +\frac{21}{10} a_{21} \delta^{\, \prime} + \frac{11}{5} a_{22} \delta^{\,\prime \,2} \right)  \, ,  
\end{equation} 
which is also solved perturbatively as an expansion 
\begin{equation}
\label{kapexpvisco}
\kappa = \kappa_0\left(1 + \delta^{\prime} \tilde{\kappa}_1 + \delta^{\prime \, 2} \tilde{\kappa}_2 +\cdots \right)\, .
\end{equation}
The result is found by the Maple program

\newprog 
{\scriptsize 
\begin{Verbatim}
DefDimension(3):
DefJ():
ExprKAPPA:=(c2pDOTc2p-(DIM+2)/2)*(c1aDOTc2p*c1aDOTc1a+c2aDOTc2p*c2aDOTc2a
 -c2pDOTc2p*c2pDOTc2p-c1pDOTc2p*c1pDOTc1p)*(1+a2*S(2,c1p)):
OmegaKAPPA:=getJexpr(1,ExprKAPPA,2):
OmegaKAPPA:=convert(OmegaKAPPA,polynom):
Expr:=Delta2*(-1/2*(1+a2*(S(2,c1p)+S(2,c2p))+a2^2*S(2,c1p)*S(2,c2p))):
mu2:=getJexpr(1,Expr,2):
Expr:=Delta4*(-1/2*(1+a2*(S(2,c1p)+S(2,c2p))+a2^2*S(2,c1p)*S(2,c2p))):
mu4:=getJexpr(1,Expr,2):
mu2:=convert(mu2,polynom):
tmpA2:={solve(mu4=mu2/DIM*4*DIM/4*(DIM+2)*(1+a2),a2)}:
a2:=tmpA2[1]:
a2:=convert(a2,polynom):
mu2:=convert(mu2,polynom):
xi0:=2*sigma^(DIM-1)/DIM*n*mu2*sqrt(2*T/m):
xi01:=coeff(xi0,dprime,1)/(2/DIM*sqrt(2*T/m)*sigma^(DIM-1)*n):
xi02:=-coeff(xi0,dprime,2)/(2/DIM*sqrt(2*T/m)*sigma^(DIM-1)*n):
Tdxi0dT:=2/DIM*sqrt(2*T/m)*sigma^(DIM-1)*n*(3/5*xi01*dprime-7/10*xi02*dprime^2):
a21:=coeff(a2,dprime,1):
a22:=coeff(a2,dprime,2):
Tda2dT:=1/10*a21*dprime+1/5*a22*dprime^2:
OmegaKAPPA:=convert(taylor(OmegaKAPPA,dprime,3),polynom):
EqKAPPA:=DIM*(kappa*Tdxi0dT+xi0*TdkappadT+xi0*kappa)
 +4/(DIM+2)*kappa*n*sigma^(DIM-1)*sqrt(2*T/m)*OmegaKAPPA
 +(DIM+2)*DIM/4*n*2*T/m*(1+2*a2+Tda2dT):
kappa:=kappa0*(1+dprime*kappa1+dprime^2*kappa2):
TdkappadT:=kappa0*(1/2+3/5*dprime*kappa1+7/10*dprime^2*kappa2):
EqKAPPA:=taylor(EqKAPPA,dprime,3):
kappa0:=solve(coeff(EqKAPPA,dprime,0),kappa0);
kappa1:=solve(coeff(EqKAPPA,dprime,1),kappa1);
kappa2:=solve(coeff(EqKAPPA,dprime,2),kappa2);
omega1:=9.28569:
omega0:=6.48562:
evalf(kappa1);
evalf(kappa2);
plot(kappa/kappa0,dprime=0..0.1,color=black,linestyle=[1]);
\end{Verbatim}
}

\noindent
Thus we found the coefficient of thermal conductivity
\begin{equation}
\label{eq:kappa_fin}
\kappa=\frac{75}{64 \sigma^2 } \sqrt{\frac{T}{\pi m}} 
\left(1 + 0.393 \delta^{\prime} + 4.904 \delta^{\prime \, 2} + \dots \right)\,.
\end{equation}
Completely analogously  we obtain the remaining transport coefficient $\mu$:
\begin{equation}
\label{eq:mu_fin1}
\mu = \frac{75}{64 \sigma^2 n } \sqrt{\frac{T^3}{\pi m}} 
\left( 1.229\, \delta^{\prime} + 1.415\, \delta^{\prime \, 2} +\dots \right) 
\end{equation} 

\section{Conclusion}
We have reported a method for the analytical evaluation of functions of the collision integral, which we call kinetic integrals, as they are encountered in the kinetic theory of Granular Gases. This method is based on computational symbolic programming, which operates with a set of known basic integrals. To illustrate the application of this method we evaluate various properties of the Granular Gases, ranging from the moments of the velocity distribution function to the transport coefficients. Two different models of the  Granular Gases are considered: The simplified model, where particles collide with a constant coefficient of restitution, and the more realistic model where the restitution coefficient depends on the impact velocity, as it follows from the collision law for viscoelastic particles. We demonstrate that evaluation of some kinetic integrals  may not in practice be done manually, while the reported method allows to obtain analytical expressions for these quantities. In the examples we generalize some previously known results for the kinetic properties of the dissipative gases.





\end{document}